\documentclass[journal,10pt,twocolumn]{IEEEtran}
\usepackage{amssymb,flushend,relsize,graphicx,amsmath,psfig,cite,float,subfigure,setspace}
\usepackage{xcolor}
\usepackage{tabularx}

\newcommand{\bi}{\begin{itemize}}
\newcommand{\ei}{\end{itemize}}

\newcommand{\be}{\begin{enumerate}}
\newcommand{\ee}{\end{enumerate}}
\newcommand{\bd}{\begin{description}}
\newcommand{\ed}{\end{description}}
\newcommand{\bc}{\begin{center}}
\newcommand{\ec}{\end{center}}
\newcommand{\bt}{\begin{tabbing}}
\newcommand{\et}{\end{tabbing}}
\newcommand{\bfig}{\begin{figure}}
\newcommand{\efig}{\end{figure}}
\newcommand{\beq}{\begin{equation}}
\newcommand{\beqarr}{\begin{eqnarray}}
\newcommand{\beqarrn}{\begin{eqnarray*}}
\newcommand{\eeq}{\end{equation}}
\newcommand{\eeqarr}{\end{eqnarray}}
\newcommand{\eeqarrn}{\end{eqnarray*}}
\newcommand{\bflr}{\begin{flushright}\vspace{-0.2in}}
\newcommand{\eflr}{\end{flushright}}
\newcommand{\bsub}{\begin{subequations}}
\newcommand{\esub}{\end{subequations}}
\newcommand{\barr}{\begin{array}}
\newcommand{\earr}{\end{array}}
\newcommand{\nn}{\nonumber}

\def\undb#1{\mbox{\bf{#1}}}

\def\dn{\stackrel{\scriptscriptstyle \triangle}{=}}

\def\arg{\mbox{arg}}

\begin{document}

\title{\huge{Optimal One- and Two-Sided Multi-level ASK Modulation\\ for RIS-Assisted Noncoherent Communication Systems}}
\author{Srijika Mukhopadhyay, Badri Ramanjaneya Reddy, Soumya P. Dash, \IEEEmembership{Senior Member, IEEE}, \\ George C. Alexandropoulos, \IEEEmembership{Senior Member, IEEE}, and Sonia~A\"{i}ssa, \IEEEmembership{Fellow, IEEE}
\thanks{S. Mukhopadhyay, B. R. Reddy, and S. P. Dash are with the School of Electrical Sciences, Indian Institute of Technology Bhubaneswar, Argul, Khordha, India (email: \{23SP06014, brr15\}@iitbbs.ac.in, soumyapdashiitbbs@gmail.com).}
\thanks{G. C. Alexandropoulos is with the Department of Informatics and Telecommunications, National and Kapodistrian University of Athens, Panepistimiopolis Ilissia, 15784 Athens, Greece and also with the Department of Electrical and Computer Engineering, University of Illinois Chicago, IL 60601, USA (email: alexandg@di.uoa.gr).}
\thanks{S. A\"{i}ssa is with the Institut National de la Recherche Scientifique (INRS), Montreal, QC, Canada (e-mail: sonia.aissa@inrs.ca).}
}
{}
\maketitle
\begin{abstract}
In this paper, we analyze the performance of one- and two-sided amplitude shift keying (ASK) modulations in single-input single-output wireless communication aided by a reconfigurable intelligent surface (RIS). Two scenarios are considered for the channel conditions: a blocked direct channel between the transmitter and the receiver, and an unblocked one. For the receiver, a noncoherent maximum likelihood detector is proposed, which detects the transmitted data signal based on statistical knowledge of the channel. The system's performance is then evaluated by deriving the symbol error probability (SEP) for both scenarios using the proposed noncoherent receiver structures. We also present a novel optimization framework to obtain the optimal one- and two-sided ASK modulation schemes that minimize the SEP under constraints on the available average transmit power for both the blocked and unblocked direct channel scenarios. Our extensive numerical investigations showcase that the considered RIS-aided communication system achieves superior error performance with both derived SEP-optimal ASK modulation schemes as compared to respective traditional ASK modulation. It is also demonstrated that, between the two proposed modulation schemes, the two-sided one yields the best SEP. The error performance is further analyzed for different system parameters, providing a comprehensive performance investigation of RIS-assisted noncoherent wireless communication systems.
\end{abstract}
\IEEEpeerreviewmaketitle
\section{Introduction}
Over the recent years, the extensive research on wireless communication systems is paving the way for the development of the sixth generation (6G) of wireless networks, which is expected to address ultra-demanding use cases, such as holographic-type communications, extended reality, tactile Internet, enhanced on-board communications, integrated sensing and communications, and global ubiquitous connectivity \cite{JiHa21,HMIMO,DT_RIS,DaJoSa22}. These visionary applications will fuel an enormous increase in data traffic and complexity, necessitating significant energy and hardware expenditures \cite{WaYo23,DISAC}. Various physical-layer technologies, including massive multiple-input multiple-output (MIMO), millimeter-wave and THz communications, and ultra-dense network architecture, have been investigated to address the envisioned requirements, some of them even in the framework of fifth-generation (5G) wireless networks. However, the network energy consumption and the hardware footprint remain critical issues.

One of the most prominent technologies for 6G wireless systems, with a notable potential to handle blockage issues and improve coverage in a cost-effective and energy-efficient manner, is the reconfigurable intelligent surface (RIS) technology \cite{RIS_multi,Ja24,RIS_tutorial,RIS_Marconi,DaKa24,DaJoAi22, BhAiDa23, VaDaAc23}. A RIS consists of programmable metasurfaces embedded with miniature elements that can manipulate electromagnetic waves intelligently, boosting signal strength at the receiver and optimizing communication links. By adjusting the phases of their impinging signals, RISs create adaptive wireless environments that can redirect these signals to desired destinations \cite{RIS_deployment,RIS_pervasive}. This inherent capability of the RIS technology is being optimized to mitigate path loss, interference, and shadowing issues from low up to THz frequencies~\cite{RIS_THz}, to enhance network coverage~\cite{MeZh23}, as well as to boost device localization~\cite{RIS_localization}, or even enable it in access-point-free scenarios, and to offer energy-efficient physical-layer security~\cite{RIS_secrecy}. RISs have also been recently integrated with unmanned aerial vehicles to enhance energy and spectrum efficiency \cite{GeDo20, MaMu21, MoAiICC20, MoAiTWC23}, used in wireless transfer and energy harvesting in different frequency bands~\cite{MEH, MoAiPIMRC22}, as well as in index modulated systems \cite{BaDaGh24, BaDaKa24, BaDaKaMa24, BaDaJoGh23} and free-space optical communication systems~\cite{NaSc21}.

Recently, the performance analysis of RIS-assisted wireless systems has been the subject of various research works. In \cite{VeMa22}, the performance of a dual-hop free-space optical radio frequency communication system aided by a RIS under phase error was investigated in terms of the outage probability and packet error rate. A joint reflecting and precoding scheme for RIS-assisted MIMO communication was proposed in \cite{YeGu20} aiming to reduce the symbol error rate. In \cite{PaDaAi24}, a RIS-based system was proposed to enhance the coverage of hybrid visible light communications in an indoor environment. A RIS-assisted non-orthogonal multiple access scheme was proposed in \cite{SuJi22} to address cases where the direct link between the base station and a user is blocked. It was shown that the transmit power decreases linearly with the number of base station antennas and quadratically with the number of RIS unit elements. The performance of RIS-empowered systems subject to Nakagami-$m$ fading was analyzed in~\cite{RIS_Nak} for one random and one coherent phase configuration schemes. In \cite{YaXi21}, a deep reinforcement learning method to jointly optimize the RIS phase shifts and the transmit precoding for a multi-user setup was studied while considering various quality-of-service constraints. In \cite{AjDaRa23, AjDaRa22}, the performance of RIS-based communication over correlated Nakagami-$m$ channels was analyzed, showing that the system's performance improves with an increase in the number of RIS elements but degrades with higher channel correlation. A closed-form upper-bound and a location-dependent expression for the ergodic capacity of RIS-empowered communication over cascaded Rician fading channel were derived in~\cite{RIS_rotation}. In \cite{WuZh19}, a joint active and passive beamforming algorithm utilizing instantaneous channel state information (CSI) was proposed for RIS-aided wireless communication. It was demonstrated that the power scaling relative to the number of RIS's reflecting elements follows a squared law.

The aforementioned works often assume the availability of perfect CSI, which is typically cumbersome due to the constraints imposed by the limited capabilities of almost passive and solely reflective RISs. Channel estimation in RIS-assisted systems is actually a challenging task~\cite{RIS_tutorial}. To circumvent the need for the usually deployed pilot-assisted channel estimation methods, noncoherent communication schemes have been proposed~\cite{GoYa19}. Such schemes eliminate the requirement for pilot signal transmission and simplify the receiver design since channel or phase estimation and compensation are unnecessary. Ongoing research on the design of optimal constellation schemes for noncoherent communications aims to further enhance energy efficiency and reduce hardware complexity \cite{ZhZh18, DuNg23, LiZh19, MaCh16, DaMaPa22, MaMu16, ReDa24, ReDaLi24}. For instance, an energy-based noncoherent constellation that depends only on statistical CSI was presented in \cite{MaCh16}, demonstrating asymptotic optimality in terms of symbol error rate. Similarly, in \cite{DaMaPa22}, the authors analyzed the performance of a RIS-assisted system with index modulation schemes using a greedy detector that does not require CSI at the receiver. In \cite{MaMu16}, optimal multi-level amplitude shift keying (ASK) for a noncoherent receive diversity system was obtained for different fading conditions, namely, uncorrelated non-identical, exponentially correlated~\cite{Exponen_Corr}, and uniformly correlated fading. Optimal multi-level one-sided ASK for a noncoherent receive diversity system subject to Rician fading was designed in~\cite{ReDa24}. In \cite{ReDaLi24}, it was demonstrated that optimal two-sided ASK outperforms one-sided ASK for a noncoherent system.

The advantages offered by noncoherent communications have recently motivated the design of noncoherent schemes for RIS-assisted wireless systems~\cite{Se22, Kun_1, Kun_2, CaHuBa23, CaHuAn23, InWa24}. It was shown in~\cite{Se22} that the performance of RIS-assisted noncoherent MIMO communication improves when the number of RIS elements increases. A noncoherent demodulation scheme for differential modulation, which was combined with a codebook-based beam training of the RIS for zero-overhead training, was presented in~\cite{Kun_1}. A RIS-assisted orthogonal frequency division multiplexing communication system, based on differential phase shift keying combined with random phase configurations at the RIS, was presented and  analyzed in~\cite{Kun_2}, which showed that, in various mobility and spatial correlation scenarios, the noncoherent system outperforms coherent demodulation. In \cite{CaHuBa23}, a noncoherent RIS-aided system using joint index keying and $M$-ary differential chaos shift keying was presented, achieving comparable error performance to coherent RIS-based spatial modulation systems. Significant improvements in error performance for a code-index-modulation-aided differential chaos shift keying system when incorporating a RIS were reported in~\cite{CaHuAn23}. Very recently, \cite{InWa24} presented a noncoherent RIS system with differential modulation, which was shown to require low training complexity while exhibiting robustness against channel variations.

Motivated by the aforementioned promising advances of noncoherent detection in RIS-aided wireless communications, in this paper we analyze the performance of RIS-aided noncoherent single-input single-output (SISO) communication systems with one- and two-sided ASK, which are modulation schemes known for their effectiveness with noncoherent detection. Two propagation scenarios are considered: the first when the direct link between the source and the destination is blocked, and the second when it is unblocked. The contributions of our work are summarized as follows:
\begin{enumerate}
    \item Optimal noncoherent maximum likelihood (ML) receiver structures for blocked and unblocked direct channel scenarios are proposed.
    \item Utilizing the said reception structures, analytical expressions for the union-bound on the SEP in both propagation scenarios are obtained.
    \item A novel optimization framework is presented to obtain the optimal one- and two-sided ASK modulation schemes for SEP minimization under transmit energy constraints for the considered propagation scenarios.
    \item Extensive numerical results are presented to highlight the performance superiority of RIS-aided SISO communications utilizing the proposed optimal one- and two-sided ASK modulation schemes in comparison to conventional one- and two-sided ASK.
\end{enumerate}
The remainder of the paper is organized as follows. Section II describes the system model and presents the proposed receiver structures for the blocked and unblocked direct channel scenarios. Using these receiver structures, section III provides the derivation of analytical expressions for the union-bound on the SEP in both scenarios. Section IV presents the optimization framework for determining the optimal ASK modulation schemes minimizing the SEP performance under average transmit energy constraint. The numerical evaluations in section V validate the SEP analysis and demonstrate the structures of the optimal modulation schemes. Finally, section VI presents the concluding remarks of the paper.
\section{System Model and Receiver Design}
A RIS-assisted wireless communication system comprising one transmitter and one receiver employing respectively multi-level ASK modulation for data transmission and symbol-by-symbol detection for noncoherent reception is considered. The communication ends are assumed to be equipped with single antennas, and the RIS consists of $L$ unit elements of continuously tunable responses. Denoting the transmitted symbol by $s$, the $L \times 1$ channel gain vector between the transmitter and the RIS by $\undb{h}_1$, the $L \times 1$ channel gain vector between the RIS and the receiver by $\undb{h}_2$, and the direct channel between the transmitter and the receiver by $h_d$, the baseband received signal can be mathematically expressed as
\beq
r = \left( \undb{h}_2^H \mathbf{\Phi} \undb{h}_1 + h_d \right) s + n \, ,
\label{eq1}
\eeq
where $\left( \cdot \right)^H$ denotes the Hermitian (complex conjugate) operator, and $\mathbf{\Phi}$ is an $L \times L$ diagonal matrix with its diagonal entries being the phase shifts introduced by the reflecting elements of the RIS. Owing to the practical implications of the deployment of a RIS to facilitate the realization of a virtual line-of-sight communication link for the transceiver pair in cases of original absence of such a link between the transmitter and the receiver, the channel gains between the transmitter-RIS and the RIS-receiver pairs are assumed to follow complex Gaussian distributions with non-zero means, implying that $\undb{h}_1 \sim {\mathcal{CN}} \left( \mu_1 \undb{1}_L , \sigma_h^2 \undb{I}_L \right)$ and $\undb{h}_2 \sim {\mathcal{CN}}\left( \mu_2 \undb{1}_L , \sigma_h^2 \undb{I}_L \right)$, where $\undb{1}_L$ and $\undb{I}_L$ denote the $L \times 1$ vector of ones and the $L \times L$ ($L\geq2$) identity matrix, respectively. Furthermore, the direct channel gain $h_d$ follows a zero-mean complex Gaussian distribution with variance $\sigma_{hd}^2$, i.e., $h_d \sim {\mathcal{CN}} \left(0,\sigma_{hd}^2 \right)$.

By assuming that the transmitter is capable of possessing the phases of the compounded channel $\undb{h}_2^H \undb{h}_1$, e.g., via a dedicated uplink channel estimation process~\cite{RIS_tutorial}, the diagonal elements of the phase configuration $\mathbf{\Phi}$ can be set such that the instantaneous power of the signal component arising due to the compounded channel is maximized. Using the notations $h_{1,\ell} \triangleq \left| h_{1,\ell} \right| e^{\jmath \angle h_{1,\ell}}$ and $h_{2,\ell} \triangleq \left| h_{2,\ell} \right| e^{\jmath \angle h_{2,\ell}}$, where $\jmath\triangleq\sqrt{-1}$ is the imaginary unit, the corresponding $\ell$-th (with $\ell=1,\ldots,L$) diagonal element of the phase-shift matrix is set as $\left( \mathbf{\Phi} \right)_{\ell,\ell} = e^{\jmath \left(\angle h_{2,\ell} - \angle h_{1,\ell} \right)}$. This implies that the received signal in (\ref{eq1}) can be expressed as
\beq
r = \left( \sum_{\ell=1}^L \left| h_{1,\ell} \right| \left| h_{2,\ell} \right| + h_d \right) s + n \, .
\label{eq2}
\eeq
Owing to the statistics of the channel gain vectors $\undb{h}_1$ and $\undb{h}_2$ and the practical consideration that RISs will consist of large numbers of unit elements~\cite{RIS_deployment}, the statistical properties of the random variable $\sum_{\ell=1}^L \left| h_{1,\ell} \right| \left| h_{2,\ell} \right|$ can be obtained using the central limit theorem (CLT) as
\beqarr
\sum_{\ell=1}^L \left| h_{1,\ell} \right| \left| h_{2,\ell} \right|
\sim {\mathcal{N}} \left( \mu_f , \sigma_f^2 \right) \, ,
\label{eq3}
\eeqarr
with the first and second-order moments given respectively by
\bsub
\beqarr
\mu_f = \alpha \sigma_h^2 \, , \quad
\sigma_f^2 = \beta \sigma_h^4 \, ,
\label{eq4a}
\eeqarr
where parameters $\alpha$ and $\beta$ are defined as
\beqarr
\alpha \! \! \! \! &\triangleq& \! \! \! \! \frac{L \pi}{4}
L_{1/2} \left( - K_1 \right) L_{1/2} \left( - K_2 \right) \, , \nn \\
\beta \! \! \! \! &\triangleq& \! \! \! \! L
\! \left[ \left(1+K_1 \right) \left(1+K_2 \right)
\! - \! \frac{\pi^2}{16} L_{1/2}^2 \! \left(-K_1 \right)
L_{1/2}^2 \! \left(-K_2 \right) \right] \, \nn \\
\label{eq4b}
\eeqarr
\esub
with $K_1\triangleq\left| \mu_1 \right|^2/\sigma_h^2$ and $K_2\triangleq\left| \mu_2 \right|^2/\sigma_h^2$ denoting the Rician parameters of the transmitter-RIS and RIS-receiver channels, respectively, and $L_{1/2} \left( \cdot \right)$ being a Laguerre polynomial \cite{GrRy07}. In the special case of Rayleigh fading channels, the expressions for $\alpha$ and $\beta$ parameters are simplified as
\beqarr
\alpha \big|_{\mu_1=\mu_2=0} = \frac{L \pi}{4} \ , \
\beta \big|_{\mu_1=\mu_2=0} =
\frac{L \left( 16-\pi^2 \right)}{16} \, .
\label{eq5}
\eeqarr
The additive noise $n$ appearing in (\ref{eq2}) follows a zero-mean complex Gaussian distribution such that $n \sim {\mathcal{CN}} \left(0,\sigma_n^2 \right)$. Finally, the transmitted symbol $s$ is chosen from a multi-level ASK constellation. In this paper, we consider two categories of $M$-level ASK modulation schemes, namely,
\begin{enumerate}
\item {\em One-sided ASK}:
\beq
s_m =  \sqrt{E_m} ,\,\,\, m = 1,  \ldots, M,
\label{eq6a}
\eeq
\item {\em Two-sided ASK}:
\beq
s_m = \begin{cases}
 -\sqrt{E_{\frac{M}{2} - m +1}} & \text{if}\,\, 1\leq m \leq \frac{M}{2}\\
 \sqrt{E_{ m - \frac{M}{2}}} & \text{if}\,\, \frac{M}{2} < m \leq M
 \end{cases},
\label{eq6b}
\eeq
\end{enumerate}
where $E_m$ (with $m=1,\ldots,M$) denotes the energy of the $m$-th symbol in the multi-level ASK constellation. We further assume that all symbols in the constellation are equi-probable. It is noted that a multi-level ASK constellation is suitable for noncoherent communications and, without loss of generality, we consider the one-sided multiple levels in the ASK constellation to be in increasing order, i.e., $\sqrt{E_m} < \sqrt{E_{m+1}}$ $\forall m \in \left\{ 1, \ldots, M-1\right\}$. For the two-sided ASK modulation scheme, it is considered that $\sqrt{E_m} < \sqrt{E_{m+1}}$ $\forall m \in \left\{ 1, \ldots, \frac{M}{2}-1\right\}$. Therefore, the average energy of the considered multi-level ASK constellation can be computed as
\beq
E_{av} = \frac{1}{M} \sum_{m=1}^M s_m^2 \, .
\label{eq7}
\eeq

In this work, we focus on the case where the receiver is incapable of performing channel estimation prior to data symbol decoding. To this end, we consider that the receiver employs the following ML-optimal noncoherent detection rule to estimate the transmitted symbol $s$, given by
\beq
\hat{s} \triangleq \arg \max_{s} \ln f \left( r \big| s \right) \, ,
\label{eq8}
\eeq
where $f \left( r \big| s \right)$ denotes the probability density function (PDF) of the received signal $r$ conditioned on $s$. In the following subsections, we consider two practical scenarios for the RIS-assisted communication system for which we derive the optimal receiver structures.
\subsection{ML-Optimal Reception when the Direct Link is Blocked}
Let us consider that the direct channel between the transmitter and the receiver is blocked. This implies that the received symbol expression in (\ref{eq2}) modifies to
\beq
r = \left( \sum_{\ell=1}^L \left| h_{1,\ell} \right| \left| h_{2,\ell} \right| \right) s + n \, .
\label{eq9}
\eeq
It can be observed that the end-to-end channel and the signal part are real-valued. Furthermore, from the statistics of the noise scalar $n$, it holds that $\Re \left\{ n\right\} \sim {\mathcal{N}} \left(0,\sigma_n^2/2 \right)$, in which $\Re \left\{ \cdot \right\}$ represents the real-part operator. We thus propose that the receiver first considers the real part of the received signal and then employs the noncoherent rule for symbol-by-symbol detection. This, in effect, would lead to the transmitted signal being corrupted by $\Re \left\{ n\right\}$, which has lower variance as compared to $n$, and thus, would result in a better reception performance. To this end, the proposed ML-optimal noncoherent ML receiver structure is expressed as
\beqarr
\hat{s}_{bc}\! \! \! \! &=& \! \! \! \! \arg \max_{s}
\ln f \left( \Re \left\{ r \right\} \big| s \right) \nn \\
&=& \! \! \! \! \arg \min_{s} \frac{\left( \Re \left\{r \right\} - \mu_f s \right)^2}
{2 \sigma_f^2 s^2 + \sigma_n^2}
+ \frac{1}{2} \ln \left( 2 \sigma_f^2 s^2 + \sigma_n^2 \right)\! ,
\label{eq10}
\eeqarr
where $\mu_f$ and $\sigma_f^2$ are given in (\ref{eq4a}).
\subsection{ML-Optimal Reception when the Direct Link is Unblocked}
We now consider that the direct channel $h_d$ between the transmitter and the receiver is not blocked, and thus, the received symbol is expressed as in (\ref{eq2}). As in the previous case, we consider symbol detection via the real part of the received symbol. To achieve the corresponding receiver structure, we next derive the statistics of the sum of the product of the magnitudes of the channels $h_{1,\ell}$ and $h_{2,\ell}$ for $\ell=1,\ldots,L$, when summed with the channel gain $h_d$, given as
\beq
\Re \left\{ \sum_{\ell=1}^L
\left|h_{1,\ell} \right| \left|h_{2,\ell} \right| + h_d \right\}
\sim {\mathcal{N}} \left(\mu_f , \sigma_f^2 + \frac{\sigma_{hd}^2}{2} \right) \, .
\label{eq11}
\eeq
Thus, the ML-optimal noncoherent receiver is expressed as
\begin{eqnarray}
&& \! \! \! \! \! \! \! \! \! \! \! \! \!
\hat{s}_{ubc} \! = \! \arg \max_{s}
\ln \! f \! \left( \! \Re \left\{ r \right\} \! \big| \! s \right)
\! = \! \arg \min_{s} \! \frac{\left( \Re \left\{r \right\} - \mu_f s \right)^2}
{\left( 2 \sigma_f^2 + \sigma_{hd}^2 \right) s^2 + \sigma_n^2} \nn \\
&& + \frac{1}{2}
\ln \left( \left(2 \sigma_f^2 + \sigma_{hd}^2 \right) s^2 \!+\! \sigma_n^2 \right) \, .
\label{eq13}
\end{eqnarray}
This receiver structure is similar to the one in (\ref{eq10}) when replacing $2\sigma_f^2$ by $2 \sigma_f^2 + \sigma_{hd}^2$.
\section{Error Rate Performance}
In this section, we analyze the performance of the RIS-assisted communication system with the proposed receiver structures, for both scenarios of the blocked direct channel and the unblocked direct channel. The performance metric is the SEP for which we derive closed-form upper-bound expressions using the union-bound technique.
\subsection{SEP for the Case of a Blocked Direct Channel}
We utilize the receiver functionality described in (\ref{eq10}) to derive an analytic expression for the SEP. Considering the $m$-th hypothesis to be true, i.e., the symbol $s_m$ is transmitted, the pairwise probability of correct decision when another symbol $s_k$ is transmitted can be expressed as
\beqarr
P_{c_{m \rightarrow k}} \! \! \! \! &=& \! \! \! \! \Pr
\left\{ \frac{\left( \Re \left\{r \right\} - \mu_f s_m \right)^2}
{2 \sigma_f^2 s_m^2 + \sigma_n^2}
+ \frac{1}{2} \ln \left( 2 \sigma_f^2 s_m^2 + \sigma_n^2 \right) \right. \nn \\
&& \quad \left. <  \frac{\left( \Re \left\{r \right\} - \mu_f s_k \right)^2}
{2 \sigma_f^2 s_k^2 + \sigma_n^2}
+ \frac{1}{2} \ln \left( 2 \sigma_f^2 s_k^2 + \sigma_n^2 \right) \right\} , \nn \\
&& \qquad \qquad \qquad \qquad
k=1,\ldots, M \, , k \neq m \, .
\label{eq14}
\eeqarr
We next define the quantities $X_m$ and $B_m$:
\beqarr
X_m \! \! \! \! & \dn & \! \! \! \! \left( \sum_{\ell=1}^L
\left| h_{1,\ell} \right| \left| h_{2,\ell} \right| - \mu_f \right) \sqrt{E_m}
+ \Re \left\{ n \right\} \, , \nn \\
B_m \! \! \! \! & \dn & \! \! \! \! 2 \sigma_f^2 E_m + \sigma_n^2 \, , m=1,\ldots,M,
\label{eq15}
\eeqarr
which after being substituted in (\ref{eq14}) yields the expression for the pairwise probability of correct decision as
\beqarr
&& \! \! \! \! \! \! \! \! \! \! \! \! \! \! \! \!
P_{c_{m \rightarrow k}} \nn \\
&& \! \! \! \! \! \! \! \!
= \Pr \left\{ \left( B_m^{-1} - B_{k}^{-1} \right) X_m^2
-
\frac{2 \mu_f X_m \left( \sqrt{E_m} - \sqrt{E_k} \right)}{B_k} \right. \nn \\
&& \qquad \qquad \left.
- \frac{\mu_f^2 \left( \sqrt{E_m} - \sqrt{E_k} \right)^2}{B_k}
< \frac{1}{2} \ln \left( \frac{B_k}{B_m} \right) \right\} \nn
\eeqarr
\beqarr
&& \! \! \! \! \! \! \! \!
= \Pr \left\{ \left( B_{m}^{-1} - B_{k}^{-1} \right)
\left( X_m + \frac{\mu_f B_m \left(\sqrt{E_k} - \sqrt{E_m} \right)}
{\left(B_k - B_m\right)} \right)^2 \right. \nn \\
&& \qquad \qquad \quad \left. < \frac{1}{2} \ln \left( \frac{B_k}{B_m} \right)
+ \frac{\mu_f^2 \left( \sqrt{E_k} - \sqrt{E_m} \right)^2}
{\left( B_k - B_m \right)} \right\} , \nn \\
&& \qquad \qquad \qquad \qquad \qquad \quad \
k=1,\ldots, M \, , k \neq m \, .
\label{eq16}
\eeqarr

By inspecting the structures of the one- and two-sided ASK constellations shown in (\ref{eq6a}) and (\ref{eq6b}), $P_{c_{m \rightarrow k}}$ can be analytically derived for three cases: $B_k > B_m$, $B_k < B_m$, and $B_k = B_m$ when $s_k \neq s_m$. For $k > m$, using (\ref{eq16}), we have $B_k > B_m$, $\left( B_m^{-1} - B_{k}^{-1} \right) > 0$, and $\ln \left( B_k/B_m\right) > 1$. Similarly, for $k < m$, the equivalent conditions are $B_k < B_m$, $\left( B_m^{-1} - B_{k}^{-1} \right) < 0$, and $\ln \left( B_k/B_m\right) < 1$. Therefore, the expression (\ref{eq16}) can be re-written as follows for the three cases:
\bsub
\beqarr
&& \! \! \! \! \! \! \! \! \! \! \! \! \! \! \! \! \! \! \! \!
P_{c_{m \rightarrow k}}
= \Pr \left\{ Y_{m,k}^2
> \left( B_{k}^{-1} - B_{m}^{-1} \right)^{-1}
\left[ \frac{1}{2} \ln \left( \frac{B_m}{B_k} \right) \right. \right. \nn \\
&& \! \! \! \left. \left.
+ \frac{\mu_f^2 \left( \sqrt{E_m} - \sqrt{E_k} \right)^2}
{\left( B_m - B_k \right)} \right] \right\} ,
k=1,\ldots, m-1 \, ,
\label{eq17a}
\eeqarr
\beqarr
&& \! \! \! \! \! \! \! \! \! \! \! \! \! \! \! \! \! \! \! \!
P_{c_{m \rightarrow k}}
= \Pr \left\{ Y_{m,k}^2
< \left( B_{m}^{-1} - B_{k}^{-1} \right)^{-1}
\left[ \frac{1}{2} \ln \left( \frac{B_k}{B_m} \right) \right. \right. \nn \\
&& \! \! \! \left. \left.
+ \frac{\mu_f^2 \left( \sqrt{E_k} - \sqrt{E_m} \right)^2}
{\left( B_k - B_m \right)} \right] \right\} , k=m+1,\ldots, M,
\label{eq17b}
\eeqarr
%
\beq
P_{c_{m \rightarrow k} }
= \Pr \left\{ \Re \left\{r \right\}
< 0 \right\} ,\,\, k =M-m+1 \, ,
\label{eq17c}
\eeq
where we have used the notation
\beqarr
&& \! \! \! \! \! \! \! \! \! \! \! \! \! \! \!
Y_{m,k} \nn \\
&& \! \! \! \! \! \! \! \! \! \! \! \!
\dn X_m + \frac{\mu_f B_m \left(\sqrt{E_k} - \sqrt{E_m} \right)}
{\left(B_k - B_m\right)} \, , \quad k (\neq m)=1,\ldots, M. \nn \\
\label{eq17d}
\eeqarr
\esub

From the statistics of the composite channel given in (\ref{eq3}) and the noise, the distribution of the random variable $Y_{m,k}$ is
\beqarr
&& \! \! \! \! \! \! \! \! \! \! \! \! \! \! \! \! \! \! \! \!
Y_{m,k} \sim {\mathcal{N}} \left( \frac{\mu_f B_m \left(\sqrt{E_m}-\sqrt{E_k} \right)}
{\left( B_m - B_k \right)} , \frac{B_m}{2} \right) \, , \nn \\
&& \qquad \qquad \qquad \qquad k=1,\ldots, M \, , k \neq m \, .
\label{eq18}
\eeqarr
Further, from (\ref{eq2}) and (\ref{eq3}), the distribution of $ \Re \left\{r \right\}$ is given by
\beqarr
\Re \left\{r \right\} \sim {\mathcal{N}} \left(\mu_f s, \sigma_f^2 s^2 + \frac{\sigma_n^2}{2} \right) .
\label{eq19}
\eeqarr
Clearly from (\ref{eq18}), $Y_{m,k}$ follows a non-zero real Gaussian distribution, implying that $Y_{m,k}^2$ is a non-central chi-squared random variable with a single degree of freedom. Therefore, utilizing the expression for the cumulative distribution function (CDF) of $Y_{m,k}^2$, the pairwise error probabilities (PEPs) in (\ref{eq17a}), (\ref{eq17b}) and (\ref{eq17c}) can be compactly expressed as
\bsub
\beqarr
&& \! \! \! \! \! \! \! \! \! \! \! \!
P_{e_{m \rightarrow k}} \nn \\
&& \! \! \! \! \! \! \! \! \! \! \!
=\!1\!-\! Q_{1/2} \left( \frac{\mu_f \sqrt{2B_m} \left(\sqrt{E_m}-\sqrt{E_k} \right)}
{ \left( B_m - B_k \right)} ,
\left( B_m B_k^{-1} - 1 \right)^{-1/2} \right. \nn \\
&& \left. \qquad \quad \times \left[ \ln \left( \frac{B_m}{B_k} \right)
+ \frac{2 \mu_f^2 \left( \sqrt{E_m} - \sqrt{E_k} \right)^2}
{\left( B_m - B_k \right)} \right]^{1/2} \right) , \nn \\
&& \qquad \qquad \qquad \qquad \qquad \qquad \quad k=1,\ldots, m-1,
\label{eq20a}
\eeqarr
\beqarr
&& \! \! \! \! \! \! \! \! \! \! \! \!
P_{e_{m \rightarrow k}} \nn \\
&& \! \! \! \! \! \! \! \! \!
= Q_{1/2} \left( \frac{\mu_f \sqrt{2B_m} \left(\sqrt{E_k}-\sqrt{E_m} \right)}
{ \left( B_k - B_m \right)} ,
\left( 1- B_m B_k^{-1} \right)^{-1/2} \right. \nn \\
&& \left. \qquad \quad \times \left[ \ln \left( \frac{B_k}{B_m} \right)
+ \frac{2 \mu_f^2 \left( \sqrt{E_k} - \sqrt{E_m} \right)^2}
{\left( B_k - B_m \right)} \right]^{1/2} \right) , \nn \\
&& \qquad \qquad \qquad \qquad \qquad \qquad \ \ k=m+1,\ldots,M,
\label{eq20b}
\eeqarr
\beqarr
&& \! \! \! \! \! \! \! \! \! \! \! \! \! \! \!
P_{e_{m \rightarrow k}} = Q \left( \frac{\mu_f s}{ \sqrt{\sigma_f^2 s^2 + \frac{\sigma_n^2}{2}}} \right)\!,\,\, k=M-m+1 ,
\label{eq20c}
\eeqarr
\esub
where $Q_{1/2} \left( \cdot, \cdot \right)$ is the Marcum-$Q$ function and $Q(\cdot)$ is the Gaussian-$Q$ function \cite{Pr83}. Using the PEP expressions and the union-bound technique, an upper bound on the SEP for the considered RIS-assisted system is derived as
\beqarr
P_e \leq \frac{1}{M}\sum_{m=1}^M
\!\!\left[\! \sum_{\substack{k=1,\\k<m}}^{M} P_{e_{m \rightarrow k}}
+ \!\!\sum_{\substack{k=1,\\k>m}}^{M} P_{e_{m \rightarrow k}} +\!\!\!\!\!\! \sum_{\substack{k =1, \\ k = M-m+1} }^{M} \!\!\!\!\!\!\!\!P_{e_{m \rightarrow k} } \right]\!\!.
\label{eq21}
\eeqarr
Upon substituting (\ref{eq20a}), (\ref{eq20b}) and (\ref{eq20c}) in (\ref{eq21}), followed by some algebraic simplifications, a closed-form upper-bound expression for the SEP of the considered system is obtained as shown in (\ref{eq22}), where $\alpha$ and $\beta$ are given in (\ref{eq4b}) for Rician faded channels and in (\ref{eq5}) for Rayleigh faded channels. In addition, $\Gamma_m$ denotes the SNR for each $m$-th symbol and is given by
\begin{figure*}[t!]
\beqarr
P_{e, \text{bc}} \leq
\frac{M-1}{2} \! \! \! \! &-& \! \! \! \! \frac{1}{M} \sum_{m=1}^M \left[ \sum_{\substack{k=1,\\k<m}}^{M}
Q_{1/2} \left( \frac{\alpha \sqrt{4 \beta \Gamma_m + \left( 2\beta+\alpha^2 \right)} }
{2 \beta \left( \sqrt{\Gamma_m} + \sqrt{\Gamma_k} \right)} , \right. \right. \nn \\
&& \! \! \! \! \! \! \! \! \! \! \! \! \! \! \! \! \! \! \! \! \! \! \! \! \! \! \! \! \! \! \! \! \! \! \! \! \! \! \! \! \! \! \!
\left.
\sqrt{\left( \frac{2 \beta \Gamma_k + (2\beta+\alpha^2)}{2 \beta \left( \Gamma_m - \Gamma_k \right)} \right)
\left( \ln \left( \frac{2 \beta \Gamma_m + (2\beta+\alpha^2)}{2 \beta \Gamma_k + \left(2\beta+\alpha^2 \right)} \right)
+ \frac{\alpha^2 \left( \sqrt{\Gamma_m} - \sqrt{\Gamma_k} \right)^2}
{\beta \left( \Gamma_m - \Gamma_k\right)} \right)} \right)
- \! \! \left. \sum_{\substack{k=1,\\k>m}}^{M}
Q_{1/2} \left( \frac{\alpha \sqrt{4 \beta \Gamma_m + \left(2\beta+\alpha^2 \right)} }
{2 \beta \left( \sqrt{\Gamma_k} + \sqrt{\Gamma_m} \right)} , \right. \right. \nn \\
&& \! \! \! \! \! \! \! \! \! \! \! \! \! \! \! \! \! \! \! \! \! \! \! \! \! \! \! \! \! \! \! \! \! \! \! \! \! \! \! \! \! \! \!
\left. \left.
\sqrt{\left( \frac{2 \beta \Gamma_k + \left(2\beta+\alpha^2 \right)}{2 \beta \left( \Gamma_k - \Gamma_m \right)} \right)
\left( \ln \left( \frac{2 \beta \Gamma_k + \left(2\beta+\alpha^2 \right)}{2 \beta \Gamma_m + \left(2\beta+\alpha^2 \right)} \right)
+ \frac{\alpha^2 \left( \sqrt{\Gamma_k} - \sqrt{\Gamma_m} \right)^2}
{\beta \left( \Gamma_k - \Gamma_m \right)} \right)} \right)\right.
\! \! - \! \! \! \! \! \! \! \! \! \!
\left. \sum_{\substack{k=1, \\ k=M-m+1}}^{M} \! \! \! \! \! \! \!
Q \left( \sqrt{ \frac{2 \alpha^2 \Gamma_m}{\left( 2\beta \Gamma_m + 2\beta +\alpha^2 \right)}} \right) \right]
\label{eq22}
\eeqarr
\noindent\rule{\textwidth}{.25pt}
\end{figure*}
\beq
\Gamma_m \dn \frac{(2\beta+\alpha^2) E_m \sigma_h^4}{\sigma_n^2},~m=1,\ldots, M,
\label{eq22a}
\eeq
yielding the expression for the average SNR per symbol as
\beq
\Gamma_{av} = \frac{1}{M}\sum_{m=1}^M \Gamma_m = \frac{(2\beta+\alpha^2) E_{av} \sigma_h^4}{\sigma_n^2}.
\label{eq23}
\eeq
\subsection{SEP for the Case of Unblocked Direct Channel}
We now utilize the receiver structure described by (\ref{eq13}) to analyze the SEP of the considered system with an unblocked direct channel. Considering the $m$-th hypothesis to be true, i.e., the symbol $s_m$ has been transmitted, the pairwise probability of correct decision with a symbol $s_k$ can be expressed as
\beqarr
&&\!\!\!\!\!\!\!\!\!\!\!\!\!
P_{c_{m \rightarrow k}} \nn \\
&&\!\!\!\!\!\!\!\!\!\!\!\!= \Pr \! \left\{ \! \! \frac{\left( \Re \left\{r \right\} - \mu_f s_m \right)^2}
{\left (2 \sigma_f^2 + \sigma_{hd}^2 \right) s_m^2 + \sigma_n^2}
\! + \! \frac{1}{2} \ln \left( \left (2 \sigma_f^2 + \sigma_{hd}^2 \right ) s_m^2 + \sigma_n^2 \right) \right. \nn \\
&&\!\!\!\!\! \! \! \! \!
\left. <  \frac{\left( \Re \left\{r \right\} - \mu_f s_k \right)^2}
{\left (2 \sigma_f^2 + \sigma_{hd}^2 \right ) s_k^2 + \sigma_n^2}
\! + \! \frac{1}{2} \ln \left( \left (2 \sigma_f^2 + \sigma_{hd}^2 \right ) s_k^2 + \sigma_n^2 \right) \right\}\!, \nn \\
&& \qquad \qquad \qquad \qquad \qquad \quad \,\,\,\,k=1,\ldots, M,\, k \neq m \, .
\label{eq24}
\eeqarr
By defining the following random variable and parameter,
\beqarr
\widetilde{X}_m \! \! \! \! & \dn & \! \! \! \! \left( \sum_{\ell=1}^L
\left| h_{1,\ell} \right| \left| h_{2,\ell} \right| + \Re \left \{ h_d \right \} - \mu_f \right) \sqrt{E_m}
+ \Re \left\{ n \right\} \, , \nn \\
\widetilde{B}_m \! \! \! \! & \dn & \! \! \! \! \left (2 \sigma_f^2 + \sigma_{hd}^2 \right ) E_m + \sigma_n^2 \, , m=1,\ldots,M \, ,
\label{eq25}
\eeqarr
and then using them in (\ref{eq24}), the expression for the pairwise probability of correct decision can be re-written as
\beqarr
&& \! \! \! \! \! \! \! \! \! \! \! \!
P_{c_{m \rightarrow k}}
\! \! = \! \Pr \! \left\{ \! \left( {\widetilde{B}}_m^{-1} - {\widetilde{B}}_{k}^{-1} \right) {\widetilde{X}}_m^2
- \frac{2 \mu_f {\widetilde{X}}_m \left( \sqrt{E_m} - \sqrt{E_k} \right)}{\widetilde{B}_k} \right. \nn \\
&& \qquad \qquad \qquad \left.
- \frac{\mu_f^2 \left( \sqrt{E_m} - \sqrt{E_k} \right)^2}{{\widetilde{B}}_k}
< \frac{1}{2} \ln \left( \frac{{\widetilde{B}}_k}{{\widetilde{B}}_m} \right) \right\} \nn \\
&& \! \! \! \! \! \! \! \!
= \Pr \left\{ \left( {\widetilde{B}}_{m}^{-1} - {\widetilde{B}}_{k}^{-1} \right)
\left( {\widetilde{X}}_m + \frac{\mu_f {\widetilde{B}}_m \left(\sqrt{E_k} - \sqrt{E_m} \right)}
{\left({\widetilde{B}}_k - {\widetilde{B}}_m\right)} \right)^2 \right. \nn \\
&& \qquad \qquad \quad \ \left. < \frac{1}{2} \ln \left( \frac{{\widetilde{B}}_k}{{\widetilde{B}}_m} \right)
+ \frac{\mu_f^2 \left( \sqrt{E_k} - \sqrt{E_m} \right)^2}
{\left( {\widetilde{B}}_k - {\widetilde{B}}_m \right)} \right\} , \nn \\
&& \qquad \qquad \qquad \qquad \qquad \quad \
k=1,\ldots, M \, , k \neq m \, .
\label{eq26}
\eeqarr
The expression in (\ref{eq26}) bears a resemblance to that in (\ref{eq16}). Consequently, by employing a similar solution methodology, the expressions for the PEPs can be derived for the cases of $B_k > B_m$, $B_k < B_m$, and $B_k = B_m$, for $s_k \neq s_m$, as
\bsub
\beqarr
&& \! \! \! \! \! \! \! \! \! \! \! \! \! \!
P_{e_{m \rightarrow k}}= 1 \nn \\
&& \! \! \! \! \! \! \! \! \! \! \!
- Q_{1/2} \left( \frac{\mu_f \sqrt{2{\widetilde{B}}_m} \left(\sqrt{E_m}-\sqrt{E_k} \right)}
{ \left( {\widetilde{B}}_m - {\widetilde{B}}_k \right)} ,
\left( {\widetilde{B}}_m {\widetilde{B}}_k^{-1} - 1 \right)^{-1/2} \right. \nn \\
&& \left. \times \left[ \ln \left( \frac{{\widetilde{B}}_m}{{\widetilde{B}}_k} \right)
+ \frac{2 \mu_f^2 \left( \sqrt{E_m} - \sqrt{E_k} \right)^2}
{\left( {\widetilde{B}}_m - {\widetilde{B}}_k \right)} \right]^{1/2} \right) , \nn \\
&& \qquad \qquad \qquad \qquad \qquad \qquad k=1,\ldots, m-1 ,
\label{eq27a}
\eeqarr
\beqarr
&& \! \! \! \! \! \! \! \! \! \! \! \! \!
P_{e_{m \rightarrow k}} \nn \\
&& \! \! \! \! \! \! \! \! \!
= Q_{1/2} \left( \frac{\mu_f \sqrt{2{\widetilde{B}}_m} \left(\sqrt{E_k}-\sqrt{E_m} \right)}
{ \left( {\widetilde{B}}_k - {\widetilde{B}}_m \right)} ,
\left( 1- {\widetilde{B}}_m {\widetilde{B}}_k^{-1} \right)^{-1/2} \right. \nn \\
&& \left. \qquad \quad \times \left[ \ln \left( \frac{{\widetilde{B}}_k}{{\widetilde{B}}_m} \right)
+ \frac{2 \mu_f^2 \left( \sqrt{E_k} - \sqrt{E_m} \right)^2}
{\left( {\widetilde{B}}_k - {\widetilde{B}}_m \right)} \right]^{1/2} \right) , \nn \\
&& \qquad \qquad \qquad \qquad \qquad \qquad \ k=m+1,\ldots, M ,
\label{eq27b}
\eeqarr
and
\beqarr
&& \! \! \! \! \! \! \! \! \! \! \! \! \! \! \!
P_{e_{m \rightarrow k} } = Q \left( \frac{\mu_f s}{ \sqrt{\left( \sigma_f^2 +\frac{\sigma_{hd}^2}{2} \right) s^2 + \frac{\sigma_n^2}{2}}} \right), \nn \\
&& \qquad \qquad \qquad\qquad \qquad \quad k = M-m+1 .
\label{eq27c}
\eeqarr
\esub
By using the union-bound expression provided in (\ref{eq22}) along with the PEP expressions in (\ref{eq27a}), (\ref{eq27b}), and (\ref{eq27c}), a closed-form upper-bound on the SEP performance for the unblocked direct channel case is obtained as shown in (\ref{eq28}).
\begin{figure*}[t!]
\beqarr
&& \! \! \! \! \! \! \! \! \! \! \! \! \! \! \! \! \! \! \!
P_{e, \text{ubc}}
\leq \frac{M-1}{2} - \frac{1}{M} \! \left[ \! \sum_{m=2}^M \sum_{k=1}^{m-1} \!
Q_{1/2} \! \! \left( \frac{\alpha \sqrt{2 \left(2\beta+1 \right) \widetilde{\Gamma}_m + \left( 2\beta+\alpha^2+1 \right)} }
{\left(2\beta+1 \right)\left( \sqrt{\widetilde{\Gamma}_m} + \sqrt{\widetilde{\Gamma}_k} \right)} , \right. \right. \nn \\
&& \qquad \left. \left.
\sqrt{\left( \frac{\left(2\beta+1 \right) \widetilde{\Gamma}_k + \left( 2\beta+\alpha^2+1 \right)}{\left(2 \beta+1 \right) \left( \widetilde{\Gamma}_m - \widetilde{\Gamma}_k \right)} \right)
\left( \ln \left( \frac{\left(2 \beta+1 \right) \widetilde{\Gamma}_m + \left( 2\beta+\alpha^2+1 \right))}{\left(2 \beta+1 \right) \widetilde{\Gamma}_k + \left(2\beta+\alpha^2+1 \right)} \right)
+ \frac{2 \alpha^2 \left( \sqrt{\widetilde{\Gamma}_m} - \sqrt{\widetilde{\Gamma}_k} \right)^2}
{\left(2 \beta+1 \right) \left( \widetilde{\Gamma}_m - \widetilde{\Gamma}_k\right)} \right)} \right) \right. \nn \\
&-& \! \! \! \! \! \! \left. \sum_{m=1}^{M-1} \sum_{k=m+1}^{M} \! \! \!
Q_{1/2} \left( \! \frac{\alpha \sqrt{2 \left(2 \beta+1 \right) \widetilde{\Gamma}_m + \left(2\beta+\alpha^2+1 \right)} }
{\left(2 \beta+1 \right) \left( \sqrt{\widetilde{\Gamma}_k} + \sqrt{\widetilde{\Gamma}_m} \right)} , \right. \right. \nn \\
&& \qquad \left. \left.
\sqrt{\left( \frac{\left(2 \beta+1 \right) \widetilde{\Gamma}_k + \left(2\beta+\alpha^2+1 \right)}{\left(2 \beta+1 \right) \left( \widetilde{\Gamma}_k - \widetilde{\Gamma}_m \right)} \right)
\left( \ln \left( \frac{\left(2 \beta+1 \right) \widetilde{\Gamma}_k + \left(2\beta+\alpha^2+1 \right)}{\left(2 \beta+1 \right) \widetilde{\Gamma}_m + \left(2\beta+\alpha^2+1 \right)} \right)
+ \frac{2 \alpha^2 \left( \sqrt{\widetilde{\Gamma}_k} - \sqrt{\widetilde{\Gamma}_m} \right)^2}
{\left(2 \beta+1 \right) \left( \widetilde{\Gamma}_k - \widetilde{\Gamma}_m \right)} \right)} \right) \right. \nn \\
&-& \! \! \! \! \! \! \left. \sum_{\substack{m=1, \\ k=M-m+1}}^{M-1} \! \! \!
Q \left( \! \sqrt{ \frac{2 \alpha^2 \widetilde{\Gamma}_m}{\left( \left(2\beta+1 \right) \widetilde{\Gamma}_m + 2\beta +\alpha^2+1 \right)}} \right)\right]
\label{eq28}
\eeqarr
\noindent\rule{\textwidth}{.2pt}
\end{figure*}
In this expression, $\widetilde{\Gamma}_m$ denotes the SNR for each $m$-th symbol and is defined considering $\sigma_{hd}=\sigma_h^2$, as
\beq
{\widetilde{\Gamma}}_m \dn \frac{\left(2\beta+\alpha^2+1 \right) E_m \sigma_h^4}{\sigma_n^2}.
\label{eq29}
\eeq
\section{Optimized $M$-Level ASK Constellations}
In this section, we present the optimal one- and two-sided ASK modulation schemes minimizing the previously derived bounds on the SEPs under an average transmit power constraint. The design optimization problem is expressed in terms of the SNR expression given in (\ref{eq23}) and (\ref{eq29}) and is then solved to find the optimal SNR values.
\subsection{Optimization for the Case of a Blocked Direct Channel}
We consider the following optimization problems for the considered one- and two-sided ASK modulation schemes for the case of a blocked direct link between the transmitter and the receiver:
\begin{enumerate}
    \item {\em One-sided ASK}:
    \bsub
    \beqarr
        &&\!\!\!\!\!\!\!\!\!\!\!\!\!\!\!\!\!\!\!\!\!\!\!\! \min_{\Gamma_1, \dots, \Gamma_M} P_{e,\, \text{bc}} \nn \\
        &&\!\!\!\!\!\!\!\!\!\!\!\!\!\!\!\!\!\!\!\!\!\!\!\! \text{s.t.} \sum_{m=1}^{M} \Gamma_m = M\Gamma_{av}\,~\text{and}~0\leq \Gamma_1 < \dots < \Gamma_M.
        \label{eq30a}
    \eeqarr
    \item {\em Two-sided ASK}:
    \beqarr
        &&\!\!\!\!\!\!\!\!\!\!\!\!\!\!\!\!\!\!\!\!\!\!\!\! \min_{\Gamma_1, \dots, \Gamma_{M/2}} P_{e,\, \text{bc}} \nn \\
        &&\!\!\!\!\!\!\!\!\!\!\!\!\!\!\!\!\!\!\!\!\!\!\!\! \text{s.t.} \sum_{m=1}^{M/2} \Gamma_m = \frac{M}{2}\Gamma_{av} ~\text{and}~0\leq \Gamma_1 < \dots < \Gamma_{M/2}.
        \label{eq30b}
    \eeqarr
    \esub
\end{enumerate}
These problems can be solved using the Lagrange multiplier technique, resulting in the following modified objective functions that need to be minimized:
\begin{enumerate}
    \item {\em One-sided ASK}:
    \bsub
    \beq
    \!\!\!\!\!\!\!\!\!\!\!\!\!\!\!\mathcal{L}_1\!\! \left(\Gamma_1,\dots,\Gamma_M,\lambda_1 \right) \!\!=\!\!P_{e,\, \text{bc}} + \lambda_1\!\left( \sum_{m=1}^{M}\Gamma_m-M\Gamma_{av}\right)\!.
    \label{eq31a}
    \eeq
    \item {\em Two-sided ASK}:
    \beq
    \!\!\!\!\!\!\!\!\!\!\!\!\!\!\!\mathcal{L}_2\!\! \left(\Gamma_1,\dots,\Gamma_{M/2},\lambda_2 \right)\!\! =\!\!P_{e,\, \text{bc}} + \lambda_2\!\left( \sum_{m=1}^{M/2}\Gamma_m-\frac{M}{2}\Gamma_{av}\right)\!.
    \label{eq31b}
    \eeq
    \esub
\end{enumerate}
In the latter expressions, $\lambda_1$ and $\lambda_2$ are the Lagrangian multipliers. The optimal SNR values, denoted by $\Gamma_{t,opt}$ for $t=1, \dots, M$, can be obtained by simultaneously solving the following equations:
\begin{enumerate}
    \item {\em One-sided ASK}:
    \bsub
    \beqarr
     \frac{\partial \mathcal{L}_1 \left(\Gamma_1,\dots,\Gamma_M,\lambda_1 \right)}{\partial \Gamma_t} \! \! \! \! &=& \! \! \! \! 0,\,\, t=1, \dots,M, \nn \\
     \frac{\partial \mathcal{L}_1 \left(\Gamma_1,\dots,\Gamma_M,\lambda_1 \right)}{\partial \lambda_1} \! \! \! \! &=& \! \! \! \! 0.
     \label{eq32a}
    \eeqarr
    \item {\em Two-sided ASK}:
    \beqarr
    \frac{\partial \mathcal{L}_2 \left(\Gamma_1,\dots,\Gamma_{M/2},\lambda_2 \right)}{\partial \Gamma_t} \! \! \! \! &=& \! \! \! \! 0,\,\, t=1, \dots,{M/2}, \nn \\
    \frac{\partial \mathcal{L}_2 \left(\Gamma_1,\dots,\Gamma_{M/2},\lambda_2 \right)}{\partial \lambda_2} \! \! \! \! &=& \! \! \! \! 0.
    \label{eq32b}
    \eeqarr
    \esub
\end{enumerate}
By substituting (\ref{eq22}) in (\ref{eq32a}) and (\ref{eq32b}), the following simplifications can be reached:
\beqarr
 &&\sum_{\substack{m=1,\\ m \neq t}}^M \frac{\partial P_{e_{m \rightarrow t} }}{\partial \Gamma_t}+ \sum_{\substack{k=1, \\k \neq t}}^{M} \frac{\partial P_{e_{t \rightarrow k}}}{\partial \Gamma_t}  + \!\!\!\! \sum_{\substack{m=1,\\k = M-m+1,\\
 m=t}}^{M-1} \frac{\partial P_{e_{m \rightarrow k}}}{\partial \Gamma_t}=0, \nn \\
&&\sum_{m=1}^{M} \Gamma_m= M\Gamma_{av} \: \text{for one-sided ASK}, \nn \\
&&\sum_{m=1}^{\frac{M}{2}} \Gamma_m= \frac{M}{2}\Gamma_{av} \: \text{for two-sided ASK}.
\label{eq33}
\eeqarr
Using the following definitions:
\beqarr
a_{ij} \! \! \! \! &\triangleq& \! \! \! \!
\frac{\alpha \sqrt{4\beta \Gamma_i+2\beta+\alpha^2}}{2\beta\left(\sqrt{\Gamma_i}+\sqrt{\Gamma_j}\right)}, \nn \\
b_{ij} \! \! \! \! &\triangleq& \! \! \! \! \sqrt{\left( \frac{2 \beta \Gamma_j +2\beta+\alpha^2}{2 \beta \left( \Gamma_i - \Gamma_j \right)} \right)} \nn \\
& \times & \! \! \! \! \sqrt{\left( \ln \left( \frac{2 \beta \Gamma_i +2\beta+\alpha^2}{2 \beta \Gamma_j + 2\beta+\alpha^2} \right)+\frac{\alpha^2 \left( \sqrt{\Gamma_i} - \sqrt{\Gamma_j} \right)^2}{\beta \left( \Gamma_i - \Gamma_j\right)} \right)}, \nn \\
c \! \! \! \! & \triangleq & \! \! \! \! \frac{2\beta+\alpha^2}{2\beta}, \text{ and} \nn \\
d_{ij} \! \! \! \! &\triangleq& \! \! \! \! \sqrt{\left( \ln \left( \frac{2 \beta \Gamma_j +2\beta+\alpha^2}{2 \beta \Gamma_i + 2\beta+\alpha^2} \right)+\frac{\alpha^2 \left( \sqrt{\Gamma_j} - \sqrt{\Gamma_i} \right)^2}{\beta \left( \Gamma_j - \Gamma_i\right)} \right)}\nn \\
& \times & \! \! \! \! \sqrt{\left( \frac{2 \beta \Gamma_j +2\beta+\alpha^2}{2 \beta \left( \Gamma_j - \Gamma_i \right)} \right)},
\label{eq34}
\eeqarr
the differential terms in (\ref{eq33}) can be re-expressed as
\bsub
\beqarr
&&\!\!\!\!\!\!\!\!\!\!\!\!\!\!\!\!\!\!\!\!\!\!\!\!\!\frac{\partial P_{e_{m \rightarrow t} \big|t<m}}{\partial \Gamma_t} \!=\! \frac{\partial}{\partial \Gamma_t} \left(1- Q_{1/2}\left(a_{mt},b_{mt} \right) \right) \nn \\
\!\!\!\!\!&=&\!\!\!\!\!\frac{\partial}{\partial \Gamma_t} \left(1- Q\left(b_{mt}-a_{mt}\right)-Q\left(b_{mt}+a_{mt} \right) \right) \nn \\
\!\!\!\!\!&=&\!\!\!\!\! -\frac{1}{\sqrt{2\pi}}\exp{\left(-\frac{\left(b_{mt}-a_{mt}\right)^2}{2} \right)} \frac{\partial \left(b_{mt}-a_{mt}\right)}{\partial \Gamma_t} \nn \\
\!\!\!\!\!&&\!\!\!\!\! - \frac{1}{\sqrt{2\pi}}\exp{\left(-\frac{\left(b_{mt}+a_{mt}\right)^2}{2} \right)} \frac{\partial \left(b_{mt}+a_{mt}\right)}{\partial \Gamma_t}, \nn \\
\label{eq35a}
\eeqarr
\beqarr
&&\!\!\!\!\!\!\!\!\!\!\!\!\!\!\!\!\!\!\!\!\!\!\!\!\!\frac{\partial P_{e_{m \rightarrow t} \big|m<t}}{\partial \Gamma_t}
= \frac{\partial}{\partial \Gamma_t} \left( Q_{1/2}\left(a_{mt},d_{mt} \right) \right) \nn \\
\!\!\!\!\!&=&\!\!\!\!\!\frac{\partial}{\partial \Gamma_t} \left(Q\left(d_{mt}-a_{mt}\right)+ Q\left(d_{mt}+a_{mt} \right) \right) \nn \\
\!\!\!\!\!&=&\!\!\!\!\! \frac{1}{\sqrt{2\pi}}\exp{\left(-\frac{\left(d_{mt}-a_{mt}\right)^2}{2} \right)} \frac{\partial \left(d_{mt}-a_{mt}\right)}{\partial \Gamma_t} \nn \\
\!\!\!\!\!&+&\!\!\!\!\! \frac{1}{\sqrt{2\pi}}\exp{\left(-\frac{\left(d_{mt}+a_{mt}\right)^2}{2} \right)} \frac{\partial \left(d_{mt}+a_{mt}\right)}{\partial \Gamma_t},
\label{eq35b}
\eeqarr
\beqarr
\frac{\partial P_{e_{t \rightarrow k} \big|k<t}}{\partial \Gamma_t}\!\!\!\!\!&=&\!\!\!\!\! \frac{\partial}{\partial \Gamma_t} \left(1- Q_{1/2}\left(a_{tk},b_{tk} \right) \right) \nn \\
&& \! \! \! \! \! \! \! \! \! \! \! \! \! \! \! \! \! \! \! \!
= \frac{\partial}{\partial \Gamma_t} \left(1- Q\left(b_{tk}-a_{tk}\right)-Q\left(b_{tk}+a_{tk} \right) \right) \nn \\
&& \! \! \! \! \! \! \! \! \! \! \! \! \! \! \! \! \! \! \! \!
= - \frac{1}{\sqrt{2\pi}}\exp{\left(-\frac{\left(b_{tk}-a_{tk}\right)^2}{2} \right)} \frac{\partial \left(b_{tk}-a_{tk}\right)}{\partial \Gamma_t} \nn \\
&& \! \! \! \! \! \! \! \! \! \! \! \! \! \! \! \! \! \! \! \! 
- \frac{1}{\sqrt{2\pi}}\exp{\left(-\frac{\left(b_{tk}+a_{tk}\right)^2}{2} \right)} \frac{\partial \left(b_{tk}+a_{tk}\right)}{\partial \Gamma_t},
\label{eq35c}
\eeqarr
\beqarr
\frac{\partial P_{e_{t \rightarrow k} \big|t<k}}{\partial \Gamma_t}\!\!\!\!\!&=&\!\!\!\!\! \frac{\partial}{\partial \Gamma_t} \left(Q_{1/2}\left(a_{tk},d_{tk} \right) \right) \nn \\
&& \! \! \! \! \! \! \! \! \! \! \! \! \! \! \! \! \! \! \! \! \! \! \! \!
= \frac{\partial}{\partial \Gamma_t} \left(Q\left(d_{tk}-a_{tk}\right)+ Q\left(d_{tk}+a_{tk} \right) \right) \nn \\
&& \! \! \! \! \! \! \! \! \! \! \! \! \! \! \! \! \! \! \! \! \! \! \! \!
= \frac{1}{\sqrt{2\pi}}\exp{\left(-\frac{\left(d_{tk}-a_{tk}\right)^2}{2} \right)} \frac{\partial \left(d_{tk}-a_{tk}\right)}{\partial \Gamma_t} \nn \\
&& \! \! \! \! \! \! \! \! \! \! \! \! \! \! \! \! \! \! \! \! \! \! \! \!
+ \frac{1}{\sqrt{2\pi}}\exp{\left( \! -\frac{\left(d_{tk}+a_{tk}\right)^2}{2} \right)} \frac{\partial \left(d_{tk}+a_{tk}\right)}{\partial \Gamma_t},
\label{eq35d}
\eeqarr
and
\beqarr
\frac{\partial P_{e_{t \rightarrow k} \big|t=M-k+1}}{\partial \Gamma_t}\!\!\!\!\!&=&\!\!\!\!\! \frac{1}{2\sqrt{2\pi}} \exp{\left(- \frac{2\alpha^2 \Gamma_t}{2\beta \Gamma_t+2\beta+\alpha^2 }\right)}\nn \\
\!\!\!\!\!&&\!\!\!\!\! \times \frac{2\alpha^2 \left(2\beta+\alpha^2 \right)}{\left(2\beta \Gamma_t+2\beta+\alpha^2 \right)^{3/2}\sqrt{2\alpha^2+\Gamma_t}}, \nn \\
\label{eq35e}
\eeqarr
\esub
where it can be easily obtained that:
\bsub
\beqarr
&&\!\!\!\!\!\!\!\!\!\!\!\!\!\!\!\!\!\!\!\!\!\!\!\!\!\!\!\!\!\!\!\!\frac{\partial a_{mt}}{\partial \Gamma_t} = - \frac{\alpha \sqrt{2\Gamma_m+c}}{2\sqrt{2\beta}\sqrt{\Gamma_t} \left(\sqrt{\Gamma_m}+\sqrt{\Gamma_t} \right)^2} , \nn \\
&&\!\!\!\!\!\!\!\!\!\!\!\!\!\!\!\!\!\!\!\!\!\!\!\!\!\!\!\!\!\!\!\!\frac{\partial a_{tk}}{\partial \Gamma_t} = \frac{\alpha\left(2\sqrt{\Gamma_k \Gamma_t}-c\right)}{2\sqrt{2\beta}\left(\sqrt{\Gamma_m}+\sqrt{\Gamma_t} \right)^2 \sqrt{\left(2\Gamma+c \right) \Gamma_t} } ,
\label{eq36a}
\eeqarr
\beqarr
\frac{\partial b_{mt}}{\partial \Gamma_t}\! \! \! \! & = & \! \! \!\! \frac{\left(\Gamma_m+c \right)}{2b_{mt}} \left[ \frac{ \ln{\left(\frac{\Gamma_m+c}{\Gamma_t+c}\right)}}{\left(\Gamma_m -\Gamma_t \right)^2}+ \frac{4 \alpha^2 \left(\sqrt{\Gamma_m}-\sqrt{\Gamma_t} \right)^2}{2\beta\left(\Gamma_m -\Gamma_t \right)^3} \right. \nn \\
\! \! \! \! & - & \! \! \!\! \left.\frac{2 \alpha^2 \left(\sqrt{\Gamma_m}+\sqrt{\Gamma_t} \right) \frac{\left(\Gamma_m -\Gamma_t \right)}{\sqrt{\Gamma_t}}}{2\beta\left(\Gamma_m -\Gamma_t \right)^3} \right] \nn \\
\! \! \! \! & + & \! \! \!\! \frac{2 \alpha^2 \left(\sqrt{\Gamma_m}-\sqrt{\Gamma_t} \right)^2-2\beta\left(\Gamma_m -\Gamma_t \right)}{2b_{mt} 2\beta\left(\Gamma_m -\Gamma_t \right)^2},
\label{eq36b}
\eeqarr
\beqarr
\frac{\partial b_{tk}}{\partial \Gamma_t}\! \! \! \! & = & \! \! \!\! \frac{\left(\Gamma_k+c \right)}{2b_{tk}} \left[ \frac{ \ln{\left(\frac{\Gamma_t+c}{\Gamma_k+c}\right)}}{\left(\Gamma_t -\Gamma_k \right)^2}+ \frac{2 \alpha^2  \frac{\left(\sqrt{\Gamma_t}-\sqrt{\Gamma_k} \right)}{\sqrt{\Gamma_t}}}{2\beta\left(\Gamma_t -\Gamma_k \right)^2} \right] \nn \\
\! \! \! \! & + & \! \! \!\! \frac{1}{2b_{tk} \left(\Gamma_t -\Gamma_k \right)}\left(\frac{\Gamma_t+c}{\Gamma_k+c}\right) ,
\label{eq36c}
\eeqarr
\beqarr
\frac{\partial d_{mt}}{\partial \Gamma_t}\! \! \! \! & = & \! \! \!\! \frac{\left(\Gamma_m+c \right)}{2d_{mt}} \left[ \frac{ \ln{\left(\frac{\Gamma_t+c}{\Gamma_m+c}\right)}}{2\beta \left(\Gamma_t -\Gamma_m \right)^2}+ \frac{2 \alpha^2  \frac{\left(\sqrt{\Gamma_t}-\sqrt{\Gamma_m} \right)}{\sqrt{\Gamma_t}}}{2\beta\left(\Gamma_t -\Gamma_m \right)^2} \right] \nn \\ \! \! \! \! & + & \! \! \!\! \frac{1}{2d_{mt} \left(\Gamma_t -\Gamma_m \right)}, \nn \\
\label{eq36d}
\eeqarr
and
\beqarr
\frac{\partial d_{tk}}{\partial \Gamma_t}\! \! \! \! & = & \! \! \!\! \frac{\left(\Gamma_k+c \right)}{2d_{tk}} \left[ \frac{ \ln{\left(\frac{\Gamma_k+c}{\Gamma_t+c}\right)}}{\left(\Gamma_k -\Gamma_t \right)^2}+ \frac{2 \alpha^2  \frac{\left(\sqrt{\Gamma_k}-\sqrt{\Gamma_t} \right)}{\sqrt{\Gamma_t}}}{2\beta\left(\Gamma_k -\Gamma_t \right)^2} \right] \nn \\
\! \! \! \! & - & \! \! \!\! \frac{1}{2d_{tk} \left(\Gamma_k -\Gamma_t \right)}\left(\frac{\Gamma_k+c}{\Gamma_t+c}\right).
\label{eq36e}
\eeqarr
\esub

By inspecting expressions (\ref{eq35a})-(\ref{eq36e}), it can be concluded that the differential terms in (\ref{eq33}) are highly nonlinear. Consequently, they can be solved numerically to determine the optimal values of $\Gamma_m$ for the one- and two-sided ASK modulation schemes under consideration.

\subsection{Optimization for the Case of Unblocked Direct Channel}
Using the Lagrange multiplier technique as before, the optimization formulations for the optimal one- and two-sided ASK modulation schemes in the case of an unblocked direct channel are expressed as follows:
\begin{enumerate}
\bsub
    \item {\em One-sided ASK}:
    \beq
    \!\!\!\!\!\!\!\!\!\!\!\!\widetilde{\mathcal{L}}_1\! \left(\!\widetilde{\Gamma}_1,\dots,\widetilde{\Gamma}_M,\widetilde{\lambda}_1 \!\right) \!=\!P_{e,\,\text{ubc}}+\widetilde{\lambda}_1\!\!\left( \sum_{m=1}^{M}\widetilde{\Gamma}_m\!-\!M\widetilde{\Gamma}_{av}\!\!\right)\!,
    \label{eq37a}
    \eeq
    \item {\em Two-sided ASK}:
    \beq
    \!\!\!\!\!\!\!\!\!\!\!\!\widetilde{\mathcal{L}}_2\! \left(\!\widetilde{\Gamma}_1,\dots,\widetilde{\Gamma}_{M/2},\widetilde{\lambda}_2 \!\right)\!=\!P_{e,\,\text{ubc}}+\widetilde{\lambda}_2\!\!\left( \sum_{m=1}^{M/2}\widetilde{\Gamma}_m\!-\!\frac{M}{2}\widetilde{\Gamma}_{av}\!\!\right)\!,
    \label{eq37b}
    \eeq
\esub
\end{enumerate}
The optimal SNR values, denoted by $\widetilde{\Gamma}_{t,opt}$ for $t=1, \dots, M$, can be obtained by simultaneously solving the equations:
\begin{enumerate}
    \item {\em One-sided ASK}:
    \bsub
    \beqarr
     \frac{\partial \widetilde{\mathcal{L}}_1 \left(\widetilde{\Gamma}_1,\dots,\widetilde{\Gamma}_M,\widetilde{\lambda}_1 \right)}{\partial \widetilde{\Gamma}_t} \! \! \! \! &=& \! \! \! \! 0,\,\, t=1, \dots,M, \nn \\
     \frac{\partial \widetilde{\mathcal{L}}_1 \left(\widetilde{\Gamma}_1,\dots,\widetilde{\Gamma}_M,\widetilde{\lambda}_1 \right)}{\partial \widetilde{\lambda}_1} \! \! \! \! &=& \! \! \! \! 0.
     \label{eq38a}
    \eeqarr
    \item {\em Two-sided ASK}:
    \beqarr
    \frac{\partial \widetilde{\mathcal{L}}_2 \left(\widetilde{\Gamma}_1,\dots,\widetilde{\Gamma}_{M/2},\widetilde{\lambda}_2 \right)}{\partial \widetilde{\Gamma}_t} \! \! \! \! &=& \! \! \! \! 0,\,\, t=1, \dots,{M/2}, \nn \\
    \frac{\partial \widetilde{\mathcal{L}}_2 \left(\widetilde{\Gamma}_1,\dots,\widetilde{\Gamma}_{M/2},\widetilde{\lambda}_2 \right)}{\partial \widetilde{\lambda}_2} \! \! \! \! &=& \! \! \! \! 0.
    \label{eq38b}
    \eeqarr
    \esub
\end{enumerate}
By substituting (\ref{eq22}) in (\ref{eq38a}) and (\ref{eq38b}), the above equations can be re-written as follows:
\beqarr
 &&\sum_{\substack{m=1,\\ m \neq t}}^M \frac{\partial P_{e_{m \rightarrow t} }}{\partial \widetilde{\Gamma}_t}+ \sum_{\substack{k=1, \\k \neq t}}^{M} \frac{\partial P_{e_{t \rightarrow k}}}{\partial \widetilde{\Gamma}_t}  + \!\!\!\! \sum_{\substack{m=1,\\k = M-m+1,\\
 m=t}}^{M-1} \frac{\partial P_{e_{m \rightarrow k}}}{\partial \widetilde{\Gamma}_t}=0, \nn \\
&&\sum_{m=1}^{M} \widetilde{\Gamma}_m= M\widetilde{\Gamma}_{av} \: \text{for one-sided ASK}, \nn \\
&&\sum_{m=1}^{\frac{M}{2}} \widetilde{\Gamma}_m= \frac{M}{2}\widetilde{\Gamma}_{av} \: \text{for two-sided ASK}.
\label{eq39}
\eeqarr
Then, by using the definitions:
\bsub
\beqarr
\widetilde{a}_{ij} \! \! \! \! &\triangleq& \! \! \! \! \frac{\alpha \sqrt{2\widetilde{\Gamma}_i+\widetilde{c}}}{\sqrt{\left(2\beta+1 \right)}\left(\sqrt{\widetilde{\Gamma}_i}+\sqrt{\widetilde{\Gamma}_j}\right)},
\eeqarr
\beqarr
\widetilde{b}_{ij} \! \! \! \! &\triangleq& \! \! \! \! \sqrt{\left( \frac{\widetilde{\Gamma}_j+\widetilde{c}}{ \left( \widetilde{\Gamma}_i - \widetilde{\Gamma}_j \right)} \right)} \nn \\
& \times & \! \! \! \! \sqrt{\left( \ln \left( \frac{\widetilde{\Gamma}_i +\widetilde{c}}{ \widetilde{\Gamma}_j + \widetilde{c}} \right)+\frac{2 \alpha^2 \left( \sqrt{\widetilde{\Gamma}_i} - \sqrt{\widetilde{\Gamma}_j} \right)^2}{\left(2\beta+1 \right) \left( \widetilde{\Gamma}_i - \widetilde{\Gamma}_j\right)} \right)}, \nn \\
\eeqarr
\beqarr
\widetilde{c} \! \! \! \! &\triangleq& \! \! \! \! \frac{2\beta+\alpha^2+1}{2\beta+1} \, , 
\eeqarr
and
\beqarr
\widetilde{d}_{ij} \! \! \! \! &\triangleq& \! \! \! \! \sqrt{\left( \ln \left( \frac{\widetilde{\Gamma}_j +\widetilde{c}}{ \widetilde{\Gamma}_i + \widetilde{c}} \right)+\frac{2 \alpha^2 \left( \sqrt{\widetilde{\Gamma}_j} - \sqrt{\widetilde{\Gamma}_i} \right)^2}{\left(2\beta+1 \right)\left( \widetilde{\Gamma}_j - \widetilde{\Gamma}_i\right)} \right)}\nn \\
& \times &\!\!\!\! \sqrt{\left( \frac{\widetilde{\Gamma}_j +\widetilde{c}}{\left( \widetilde{\Gamma}_j - \widetilde{\Gamma}_i \right)} \right)},
\label{eq40}
\eeqarr
\esub
the differential terms in (\ref{eq39}) are obtained as:
\bsub
\beqarr
&&\!\!\!\!\!\!\!\!\!\!\!\!\!\!\!\!\!\!\!\!\!\!\!\!\frac{\partial P_{e_{m \rightarrow t} \big|t<m}}{\partial \widetilde{\Gamma}_t}\!\!\!\!\nn \\
&&\!\!\!\!\!\!\!\!\!\!\!\!\!\!\!\!\!= -\frac{1}{\sqrt{2\pi}}\exp{\left(-\frac{\left(\widetilde{b}_{mt}-\widetilde{a}_{mt}\right)^2}{2} \right)} \frac{\partial \left(\widetilde{b}_{mt}-\widetilde{a}_{mt}\right)}{\partial \widetilde{\Gamma}_t} \nn \\
&&\!\!\!\!\!\!\!\!\!\!\!\! - \frac{1}{\sqrt{2\pi}}\exp{\left(-\frac{\left(\widetilde{b}_{mt}+\widetilde{a}_{mt}\right)^2}{2} \right)} \frac{\partial \left(\widetilde{b}_{mt}+\widetilde{a}_{mt}\right)}{\partial \widetilde{\Gamma}_t},
\eeqarr
\beqarr
&&\!\!\!\!\!\!\!\!\!\!\!\!\!\!\!\!\!\!\!\!\!\!\!\!\frac{\partial P_{e_{m \rightarrow t} \big|m<t}}{\partial \widetilde{\Gamma}_t} \nn \\
&&\!\!\!\!\!\!\!\!\!\!\!\!\!\!\!\!\!= \frac{1}{\sqrt{2\pi}}\exp{\left(-\frac{\left(\widetilde{d}_{mt}-\widetilde{c}_{mt}\right)^2}{2} \right)} \frac{\partial \left(\widetilde{d}_{mt}-\widetilde{c}_{mt}\right)}{\partial \widetilde{\Gamma}_t} \nn \\
&&\!\!\!\!\!\!\!\!\!\!\!\! + \frac{1}{\sqrt{2\pi}}\exp{\left(-\frac{\left(\widetilde{d}_{mt}+\widetilde{c}_{mt}\right)^2}{2} \right)} \frac{\partial \left(\widetilde{d}_{mt}+\widetilde{c}_{mt}\right)}{\partial \widetilde{\Gamma}_t},
\label{eq41a}
\eeqarr
\beqarr
&&\!\!\!\!\!\!\!\!\!\!\!\!\!\!\!\!\frac{\partial P_{e_{t \rightarrow k} \big|k<t}}{\partial \widetilde{\Gamma}_t}\nn\\
&& \! \! \! \! \! \! \! \!
= - \frac{1}{\sqrt{2\pi}}\exp{\left(-\frac{\left(\widetilde{b}_{tk}-\widetilde{a}_{tk}\right)^2}{2} \right)} \frac{\partial \left(\widetilde{b}_{tk}-\widetilde{a}_{tk}\right)}{\partial \widetilde{\Gamma}_t} \nn \\
&& \! \! \! \! - \frac{1}{\sqrt{2\pi}}\exp{\left(-\frac{\left(\widetilde{b}_{tk}+\widetilde{a}_{tk}\right)^2}{2} \right)} \frac{\partial \left(\widetilde{b}_{tk}+\widetilde{a}_{tk}\right)}{\partial \widetilde{\Gamma}_t},
\eeqarr
\beqarr
&&\!\!\!\!\!\!\!\!\!\!\!\!\!\!\!\!\frac{\partial P_{e_{t \rightarrow k} \big|t<k}}{\partial \widetilde{\Gamma}_t}\nn \\
&& \! \! \! \! \! \! \! \!
=\frac{1}{\sqrt{2\pi}}\exp{\left(-\frac{\left(\widetilde{d}_{tk}-\widetilde{c}_{tk}\right)^2}{2} \right)} \frac{\partial \left(\widetilde{d}_{tk}-\widetilde{c}_{tk}\right)}{\partial \widetilde{\Gamma}_t} \nn \\
&& \! \! \! \! + \frac{1}{\sqrt{2\pi}}\exp{\left(-\frac{\left(\widetilde{d}_{tk}+\widetilde{c}_{tk}\right)^2}{2} \right)} \frac{\partial \left(\widetilde{d}_{tk}+\widetilde{c}_{tk}\right)}{\partial \widetilde{\Gamma}_t},
\label{eq41b}
\eeqarr
and
\beqarr
&&\!\!\!\!\!\!\!\!\!\!\!\!\!\!\!\!\!\!
\frac{\partial P_{e_{t \rightarrow k} \big|t=M-k+1}}{\partial \widetilde{\Gamma}_t}\nn \\
&=&\!\!\!\!\!\! \frac{1}{2\sqrt{2\pi}} \exp{\left(- \frac{2\alpha^2 \widetilde{\Gamma}_t}{\left(2\beta+1\right) \widetilde{\Gamma}_t+2\beta+\alpha^2+1 }\right)}\nn \\
\!\!\!\!\!\!&&\!\!\!\!\!\! \times \frac{2\alpha^2 \left(2\beta+\alpha^2+1 \right)}{\left(\left(2\beta+1\right) \widetilde{\Gamma}_t+2\beta+\alpha^2+1 \right)^{3/2}\sqrt{2\alpha^2+\widetilde{\Gamma}_t}}, \nn \\
\label{eq41c}
\eeqarr
where
\esub
\bsub
\beq
\frac{\partial \widetilde{a}_{mt}}{\partial \widetilde{\Gamma}_t} = - \frac{\alpha \sqrt{2\widetilde{\Gamma}_m+\widetilde{c}}}{2\sqrt{\left(2\beta+1\right)\widetilde{\Gamma}_t} \left(\sqrt{\widetilde{\Gamma}_m}+\sqrt{\widetilde{\Gamma}_t} \right)^2} ,
\label{eq42a}
\eeq
\beq
\frac{\partial \widetilde{a}_{tk}}{\partial \widetilde{\Gamma}_t} = \frac{\alpha\left(2\sqrt{\widetilde{\Gamma}_k \widetilde{\Gamma}_t}-\widetilde{c}\right)\left(\sqrt{\widetilde{\Gamma}_m}+\sqrt{\widetilde{\Gamma}_t} \right)^{-2}}{\left(2\sqrt{2\beta+1}\right) \sqrt{\left(2\widetilde{\Gamma}+\widetilde{c} \right)\widetilde{\Gamma}_t} } ,
\label{eq42b}
\eeq
\beqarr
\frac{\partial \widetilde{b}_{mt}}{\partial\widetilde{\Gamma}_t}\! \! \! \! & = & \! \! \!\!  \left[ \frac{ \ln{\left(\frac{\widetilde{\Gamma}_m+\widetilde{c}}{\widetilde{\Gamma}_t+\widetilde{c}}\right)}}{\left(\widetilde{\Gamma}_m -\widetilde{\Gamma}_t \right)^2}+ \frac{4 \alpha^2 \left(\sqrt{\widetilde{\Gamma}_m}-\sqrt{\widetilde{\Gamma}_t} \right)^2}{\left(2\beta+1 \right) \left(\widetilde{\Gamma}_m -\widetilde{\Gamma}_t \right)^3} \right. \nn \\
\! \! \! \! & - & \! \! \!\! \left.\frac{2 \alpha^2 \left(\sqrt{\widetilde{\Gamma}_m}+\sqrt{\widetilde{\Gamma}_t} \right) \frac{\left(\widetilde{\Gamma}_m -\widetilde{\Gamma}_t \right)}{\sqrt{\widetilde{\Gamma}_t}}}{\left(2\beta+1 \right) \left(\widetilde{\Gamma}_m -\widetilde{\Gamma}_t \right)^3} \right]\frac{\left(\widetilde{\Gamma}_m+\widetilde{c} \right)}{2\widetilde{b}_{mt}} \nn \\
\! \! \! \! & + & \! \! \!\! \frac{2 \alpha^2 \left(\sqrt{\widetilde{\Gamma}_m}-\sqrt{\widetilde{\Gamma}_t} \right)^2-\left(2\beta+1 \right)\left(\widetilde{\Gamma}_m -\widetilde{\Gamma}_t \right)}{2\widetilde{b}_{mt} \left(2\beta+1 \right)\left(\widetilde{\Gamma}_m -\widetilde{\Gamma}_t \right)^2}, \nn \\
\label{eq42c}
\eeqarr
\beqarr
\frac{\partial \widetilde{b}_{tk}}{\partial \widetilde{\Gamma}_t}\! \! \! \! & = & \! \! \!\! \frac{\left(\widetilde{\Gamma}_k+\widetilde{c} \right)}{2\widetilde{b}_{tk}} \left[ \frac{ \ln{\left(\frac{\widetilde{\Gamma}_t+\widetilde{c}}{\widetilde{\Gamma}_k+\widetilde{c}}\right)}}{\left(\widetilde{\Gamma}_t -\widetilde{\Gamma}_k \right)^2}+ \frac{2 \alpha^2  \frac{\left(\sqrt{\widetilde{\Gamma}_t}-\sqrt{\widetilde{\Gamma}_k} \right)}{\sqrt{\widetilde{\Gamma}_t}}}{\left(2\beta+1 \right)\left(\widetilde{\Gamma}_t -\widetilde{\Gamma}_k \right)^2} \right] \nn \\
\! \! \! \! & + & \! \! \!\! \frac{1}{2 \widetilde{b}_{tk} \left(\widetilde{\Gamma}_t -\widetilde{\Gamma}_k \right)}\left(\frac{\widetilde{\Gamma}_t+\widetilde{c}}{\widetilde{\Gamma}_k+\widetilde{c}}\right) , \nn \\
\label{eq42d}
\eeqarr
\beqarr
\frac{\partial \widetilde{d}_{mt}}{\partial \widetilde{\Gamma}_t}\! \! \! \! & = & \! \! \!\! \frac{\left(\widetilde{\Gamma}_m+\widetilde{c} \right)}{2 \widetilde{d}_{mt}} \left[ \frac{ \ln{\left(\frac{\widetilde{\Gamma}_t+\widetilde{c}}{\widetilde{\Gamma}_m+\widetilde{c}}\right)}}{\left(2\beta+1 \right) \left(\widetilde{\Gamma}_t -\widetilde{\Gamma}_m \right)^2} \right. \nn \\ \! \! \! \! & + & \! \! \!\! \left. \frac{ \frac {2 \alpha^2 \left(\sqrt{\widetilde{\Gamma}_t}-\sqrt{\widetilde{\Gamma}_m} \right)}{\sqrt{\widetilde{\Gamma}_t}}}{\left(2\beta+1 \right)\left(\widetilde{\Gamma}_t -\widetilde{\Gamma}_m \right)^2} \right] + \frac{1}{2\widetilde{d}_{mt} \left(\widetilde{\Gamma}_t -\widetilde{\Gamma}_m \right)}, \nn \\
\label{eq42e}
\eeqarr
and
\beqarr
\frac{\partial \widetilde{d}_{tk}}{\partial \widetilde{\Gamma}_t}\! \! \! \! & = & \! \! \!\! \frac{\left(\widetilde{\Gamma}_k+\widetilde{c} \right)}{2 \widetilde{d}_{tk}} \left[ \frac{ \ln{\left(\frac{\widetilde{\Gamma}_k+\widetilde{c}}{\widetilde{\Gamma}_t+\widetilde{c}}\right)}}{\left(\widetilde{\Gamma}_k -\widetilde{\Gamma}_t \right)^2}+ \frac{\frac{2 \alpha^2\left(\sqrt{\widetilde{\Gamma}_k}-\sqrt{\widetilde{\Gamma}_t} \right)}{\sqrt{\widetilde{\Gamma}_t}}}{\left(2\beta+1 \right) \left(\widetilde{\Gamma}_k -\widetilde{\Gamma}_t \right)^2} \right] \nn \\
\! \! \! \! & - & \! \! \!\! \frac{1}{2 \widetilde{d}_{tk} \left(\widetilde{\Gamma}_k -\widetilde{\Gamma}_t \right)}\left(\frac{\widetilde{\Gamma}_k+\widetilde{c}}{\widetilde{\Gamma}_t+\widetilde{c}}\right) \, .
\label{eq42f}
\eeqarr
\esub

The terms appearing in (\ref{eq39}) needed to determine the optimal ASK constellation are highly non-linear. Therefore, we solve this problem numerically to obtain the optimal values of $\widetilde{\Gamma}_m$ for the considered one- and two-sided ASK modulations.
\section{Numerical Results and Discussion}
This section presents the numerical evaluation of the performance metrics derived for the considered RIS-assisted noncoherent SISO communication system for varying system parameters. The system's error performance is evaluated by considering the transmitter to employ the traditional equispaced one- and two-sided ASK modulation schemes as well as the optimal one- and two-sided ASK modulation schemes proposed  in this paper. Furthermore, numerical evaluations are provided for the blocked and the unblocked direct channel scenarios. The optimal one- and two-sided ASK modulation schemes are given in (\ref{eq32a}) and (\ref{eq32b}) for the case of a blocked direct channel, and in (\ref{eq38a}) and (\ref{eq38b}) for the case of an unblocked direct channel. The curves in the figures resulting from the numerical evaluation of the derived analytical expressions are marked by ``ana.'', and the ones obtained by means of Monte-Carlo simulations are indicated by the ``sim.'' mark.

\begin{figure}[!t]
    \centering
    \includegraphics[height=2.4in,width=3.2in]{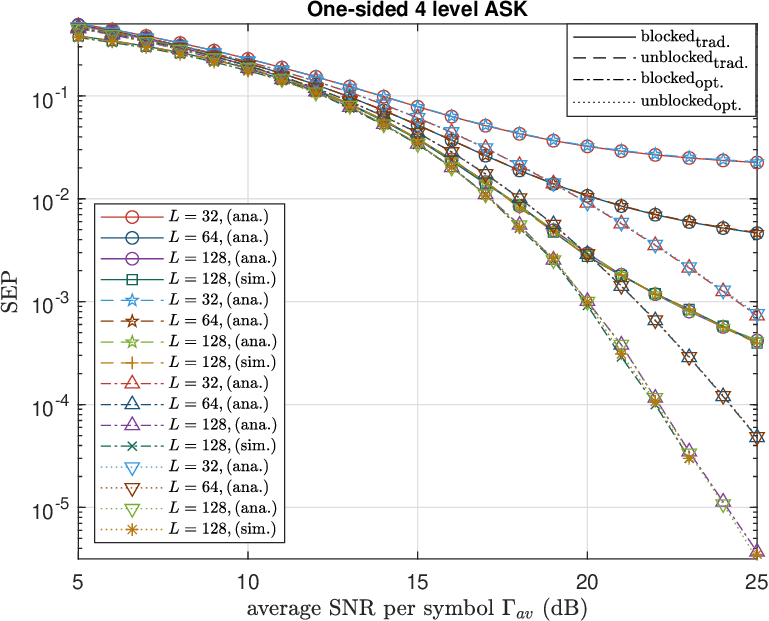}
    \caption{SEP versus average SNR per symbol with $L= 32, 64, 128$ and $M =4$ for the cases of blocked and unblocked direct channel considering one-sided ASK modulation.}
    \label{fig:1}
\end{figure}

Figure~\ref{fig:1} presents the plots for the SEP versus the average SNR per symbol $\Gamma_{av}$ for the considered RIS-assisted system with and without the direct channel employing a $4$-level {\em one-sided} ASK modulation scheme for data transmission. It can be observed that the simulation results closely resemble the analytical ones with increasing SNR, showcasing the accuracy of our union-bound approach to approximate the system's SEP. Moreover, it can be seen that the SEP values tend to saturate at higher SNR values for the system employing the traditional equispaced one-sided ASK modulation for data transmission, and this saturation is more prominent for the scenario where the direct channel between the transceiver pair is absent. In addition, SEP results for the derived optimal one-sided $4$-level ASK scheme are included in the figure, and demonstrate that this scheme leads to superior SEP performance than the traditional constellation. Interestingly, the SEP results obtained using the optimal ASK scheme do not saturate, and thus, employing this scheme improves the diversity order of the system. Furthermore, the system's performance improves with the increase of the reflecting elements of the RIS, and this improvement is boosted further with the deployment of the optimal ASK constellation. It is finally shown that, for all values of $L$, the error performance of the system is better when the direct channel is unblocked.

\begin{figure}[!t]
    \centering
    \includegraphics[height=2.4in,width=3.2in] {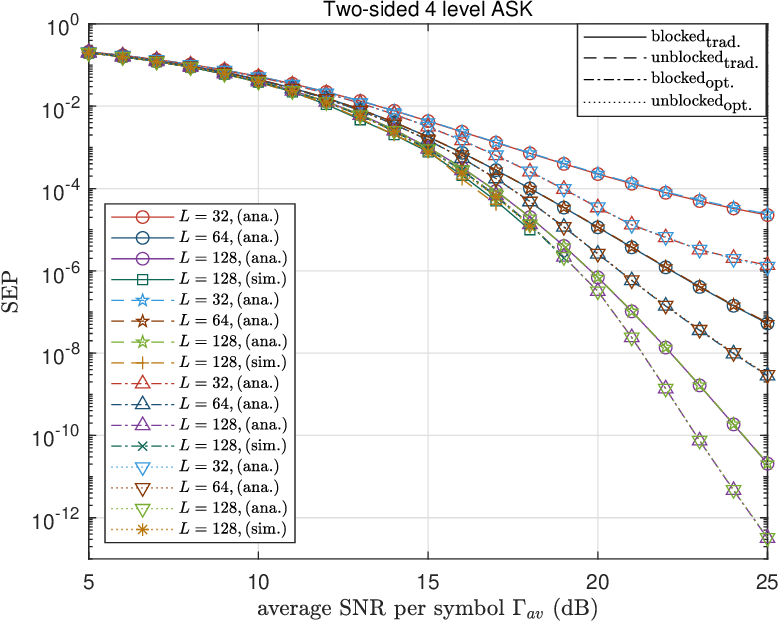}
    \caption{SEP versus the average SNR per symbol with $L=32,~64,~128$ RIS elements and $M=4$ modulation levels for both scenarios of a blocked and an unblocked direct channel considering two-sided ASK modulation.}
    \label{fig:2}
\end{figure}

In Fig.~\ref{fig:2}, we present similar investigations as in Fig.~\ref{fig:1} considering the traditional and the proposed optimal {\em two-sided} 4-level ASK modulation schemes. It is demonstrated that the SEP values tend to saturate as SNR increases, while they decrease as the number of RIS reflecting elements increases. This implies that the saturation effect is dependent on the use of the traditional equispaced ASK constellation irrespective of the one- or two-sided versions. Additionally, it can be seen that the optimal two-sided ASK constellation has better error performance than the traditional equispaced ASK constellation. In this analysis, equal constellation energy is considered for the one-sided ASK schemes used in Fig.~\ref{fig:1} as well as the two-sided ASK. It is observed that the two-sided ASK modulation scheme consistently provides better error performance than the one-sided counterpart, regardless of the SNR value and the number of reflecting elements. Similar to the observation in Fig. \ref{fig:1}, the performance of the system with unblocked direct link is superior as compared to the RIS-assisted  system with a blocked channel between the transceiver pair.

\begin{figure}[!t]
    \centering
    \includegraphics[height=2.4in,width=3.2in]{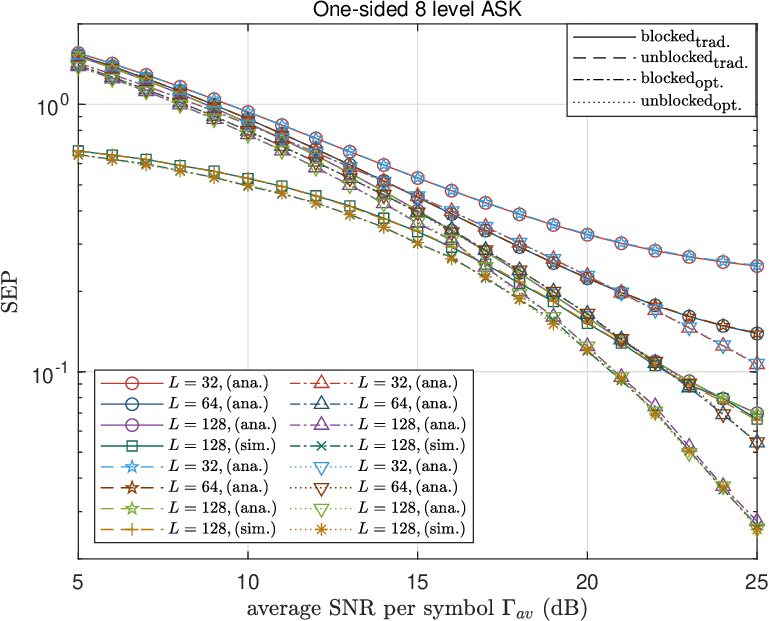}
    \caption{SEP versus the average SNR per symbol with $L= 32,~64,~128$ RIS elements and $M=8$ modulation levels for the scenarios of a blocked and an unblocked direct channel, considering one-sided ASK modulation.}
    \label{fig:3}
\end{figure}

\begin{figure}[!t]
    \centering
    \includegraphics[height=2.4in,width=3.2in] {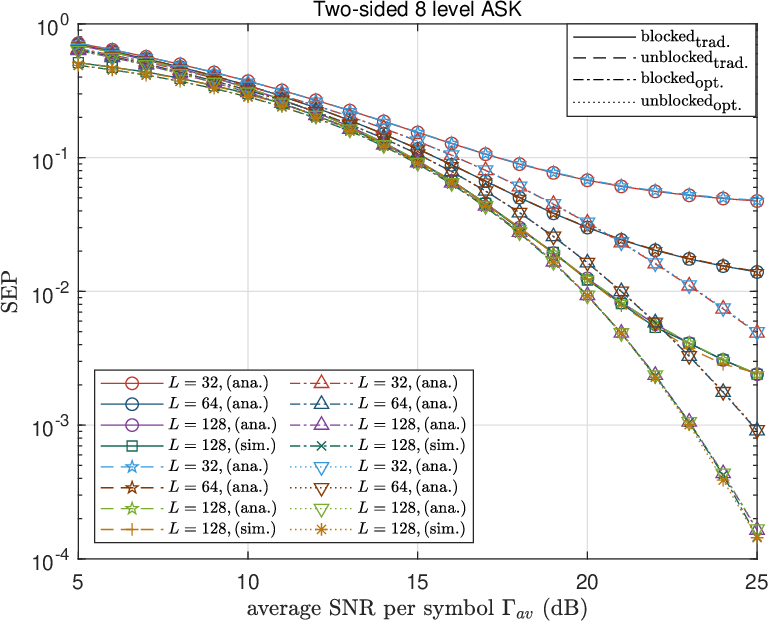}
    \caption{SEP versus the average SNR per symbol with $L= 32,~64,~128$ RIS elements and $M=8$ modulation levels for the scenarios of a blocked and an unblocked direct channel, considering two-sided ASK modulation.}
    \label{fig:4}
\end{figure}

Figures~\ref{fig:3} and~\ref{fig:4} illustrate the variation of the SEP performance with respect to the average SNR for the one- and two-sided ASK modulations, respectively, using a modulation order of $M=8$. The results show that an increase in the modulation order negatively impacts performance. This trend is consistent for both modulation schemes. Additionally, similar to the trends in Figs.~\ref{fig:1} and~\ref{fig:2}, the SEP results obtained using traditional one- and two-sided ASK modulation schemes saturate as the average SNR increases. Furthermore, better error performance is observed with optimal modulation schemes compared to traditional modulation.

\begin{figure}[!t]
    \centering
    \includegraphics[height=2.4in,width=3.2in] {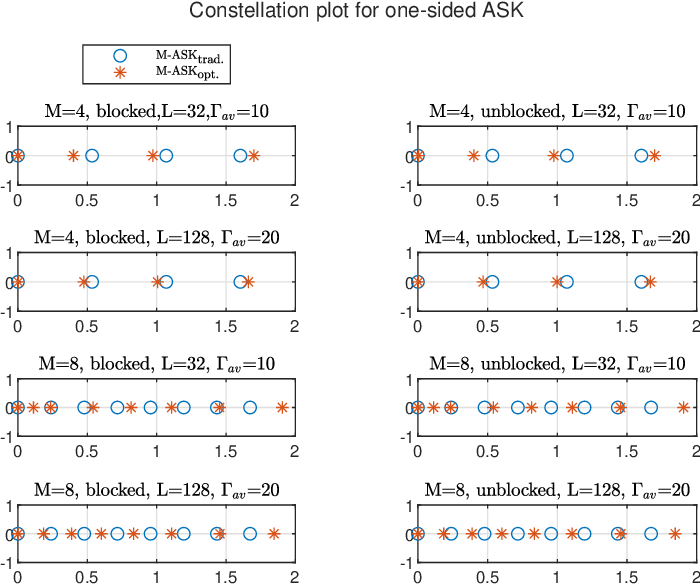}
    \caption{Optimal and traditional one-sided $M$-level ASK constellation diagrams for $M=4$ and $8$ modulation levels, $L=32$ and $128$ RIS elements, and $ \Gamma_{av}=10$ and $20$ dB.}
    \label{fig:5}
\end{figure}

\begin{figure}[!t]
    \centering
    \includegraphics[height=2.4in,width=3.2in] {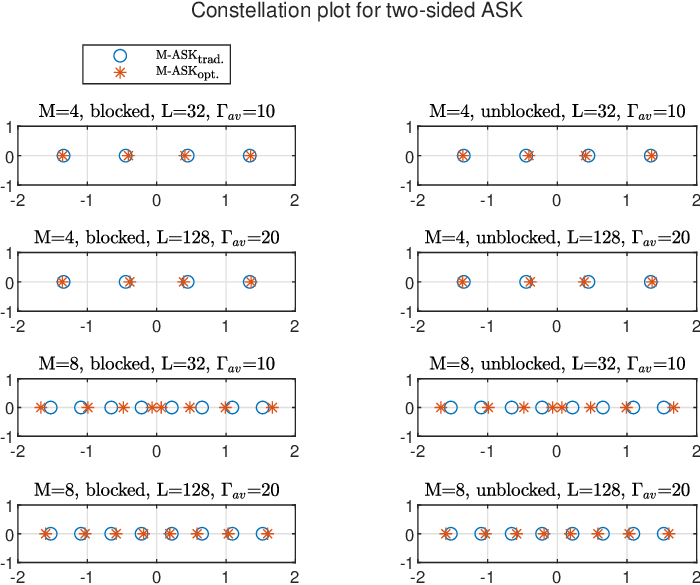}
    \caption{Optimal and traditional two-sided $M$-level ASK constellation diagrams for $M=4$ and $8$ modulation levels, $L=32$ and $128$ RIS elements, and $ \Gamma_{av}=10$ and $20$ dB.}
    \label{fig:6}
\end{figure}

Finally, in Figs.~\ref{fig:5} and~\ref{fig:6}, we compare the traditional and optimal constellation points for the one- and two-sided ASK modulation schemes, respectively, with modulation orders $M=4$ and $8$. To ensure a fair comparison, both the traditional and optimal constellations are normalized by the average SNR. It can be observed that the optimal constellation points of the one- and two-sided ASK differ from their traditional equispaced counterparts. This difference is more pronounced at lower SNR values and with fewer RIS elements. As the SNR and $L$ increase, the gap between the traditional and optimal constellation points reduces. Additionally, it is shown that the optimal constellation points for the one- and two-sided ASK are not uniformly spaced. In fact, the spacing between the optimal points increases as the energy of the constellation points increases. Moreover, the lower energy optimal constellation points in both modulation schemes are more sensitive to changes in the SNR and the number $L$.

\section{Conclusions}
This paper considered a RIS-assisted noncoherent SISO communication system where the channels between the transmitter-RIS and the RIS-receiver pairs follow the Rician distribution and presented optimal receiver structures for one- and two-sided ASK modulated data transmissions. Two cases of the direct channel between the transceiver pair were studied, namely, blocked or unblocked, with the envelope of the direct channel following the Rayleigh distribution. Capitalizing on the proposed receiver structures, closed-form expressions for the union-bound on the SEP performance were derived for both channel scenarios. The system was observed to exhibit better error rates with the two-sided modulation rather than the one-sided one, and for the case of an unblocked direct channel rather than a blocked one. An optimization framework to obtain the optimal constellation that minimizes the system's SEP under average transmit energy constraint was also proposed. It was demonstrated that the system employing the proposed ML-optimal ASK modulation schemes outperforms the traditional equispaced counterparts. The optimal constellation points of the one- and two-sided ASK modulations differ from their traditional equispaced counterparts, which were found to be dependent on the SNR and the number of the RIS unit elements.






\end{document}